\documentclass[review]{elsarticle}

\usepackage{lineno,hyperref}
\modulolinenumbers[5]

\usepackage{amsfonts}
\usepackage{amsmath}
\usepackage{amssymb}
\usepackage{graphicx}%
\usepackage{subcaption}
\usepackage{booktabs}
\begin{document}

\begin{frontmatter}

\title{Hierarchical evolutive systems, fuzzy categories and the living single cell }

\author[ifly,gamefi]{Alejandro M. Mes\'{o}n}
\ead{meson@iflysib.unlp.edu.ar}

\author[ifly,utn]{C. Manuel Carlevaro}
\ead{manuel@iflysib.unlp.edu.ar}

\author[ifly,gamefi]{Fernando Vericat\corref{mycorrespondingauthor}}
\cortext[mycorrespondingauthor]{Corresponding author}
\ead{vericat@iflysib.unlp.edu.ar}

\address[ifly]{Instituto de F\'{\i}sica de L\'{\i}quidos y Sistemas Biol\'{o}gicos
(IFLYSIB)-CONICET- CCT La Plata, Argentina.}
\address[gamefi]{Grupo de Aplicaciones Matem\'{a}ticas y Estad\'{\i}sticas de la Facultad
de Ingenier\'{\i}a (GAMEFI), Universidad Nacional de La Plata, Argentina.}
\address[utn]{Universidad Tecnol\'{o}gica Nacional - FRBA, UDB F\'{\i}sica, Mozart No
2300, C1407IVT Buenos Aires, Agentina.}

%
%
%

\begin{abstract}
In this article, the theory of hierarchical evolutive systems of Ehresmann and
Vandremeersch [Bull. Math. Bio. \textbf{49}, 13-50 (1987)] is improved by
considering the categories of the theory as fuzzy sets whose elements are the
composite objects formed by the arrows and corresponding vertices of their
embedded graphs. This way each category can be represented as a point in the
states space $\left[  0,1\right]  ^{N}$ $\subset$\ $\mathbb{R}^{N}$. The introduction of a diffeomorphism, that acts in this context as a
functor between categories, allows to define a measure preserving dynamical
system. In particular, we apply this formalism to describe a living single
cell. We propose for its state at a given time a hirerchical category with
three levels (molecular, coarse-grained and cellular levels) related by
adequate colimits. Each level involves the main functional and structural
modules in which the cell can be partitioned. The time evolution of the cell
is drived by a transformation which is a $N$-dimensional generalization of the
Ricker map whose parameters we propose to be determined by requiring that, as
hallmark of its behavior, the living cell to evolve at the edge of chaos. From
the dynamical point of view this property manifests in the fact that the
largest Lyapunov exponent is equal to zero. Since in a rather complete model
of the living cell the huge number of involved parameters can make of the
calculations a hard task, we also propose a toy model, with fewer parameters
to be determined, which emphasizes the cellular fission.
\end{abstract}


\end{frontmatter}


\section{Introduction}

The efforts made to theoretically describe complex systems have a long history
with a broad spectrum of results depending of the involved
complexity\cite{Badii1}. In the most successful extreme we have the
statistical physics of systems of many identical particles interacting among
them via relatively simple forces\cite{Gallavotti1,Honerkamp1}. But, even for
many physical systems, the difficulty of the problems is sometimes far beyond
the applicability of statistical physics methods. In disciplines such as
economics, social sciences and very particularly biological sciences, the
troubles are generally worse being the possibility of applying accurate
mathematical techniques almost null in most of the cases.

Perhaps the non physical areas in which there have been major advances in this
sense are in the computer sciences. At this respect we must call the attention
on the particular relationship between the dynamical systems theory and
aspects of computation such as the computational capability in cellular
autonoma\cite{Packard1,Mitchell1}. Of course, the Boltzmann\'{}s dynamical foundation of statistical mechanics\cite{Gallavotti1,Dorfman1} has
already been showing us, for a long time now, the potentiality of the theory
of dynamical systems to study complex systems.

In typical systems describing social or biological phenomena, we have diverse
kinds of elements with their own properties and functions as well as the
corresponding interactions among them. The elements number is not large enough
as to apply the law of large numbers to them, so the statistical tools are in
general not useful for studying these systems.

A convenient way to account for biological systems, within the philosophy of
the general systems theory\cite{Bertalanffy1}, is given by graphs, where the
vertices are the diverse elements of the system and the arrows linking them
the functional relations among these elements. This is the point of view taken
by the Chicago school\cite{Rashevsky1} in its relational theory of biological
systems\cite{Rosen1,Rosen2}. The approach appeals to the categories
theory\cite{Eilenberg1} as the main mathematical tool. However the initial
works in this direction consider just the basic notions of categories as well
as their simplest constructions.

Subsequent work focused on the response to central questions about complex
systems: the \textit{binding problem}, the\textit{ emergence problem }and the
\textit{hierarchy problem.} A.C. Ehresmann \ and J-P.Vandremeersch attempted
to give answer to these and other fundamentals problems of complex systems in
their theory of hierarchical evolutive systems\cite{Ehresmann1,Ehresmann2}. In
this theory the state of the system under study at a given time is represented
by a hierarchical category and its evolution along the time by a sequence of
such categories. The passage from the state at a given time to another state
in the sequence being governed by a functor. Each functor is determined by a
list of objectives that transform the considered category into the new state
by adding new elements, or eliminating others, or bounding previously unbound
patterns and colimits, etc. These operations are mainly descriptive, a
quantitative formulation of them being, in general, a very hard task. Thus,
the application of the modern theory of dynamical systems in order to follow
the evolution in time of complex systems, particularly biological systems,
which are described using the hierarchical evolutive theory is not as direct
as would be desirable.

In this article we try to close this gap by introducing the concept of fuzzy
categories in the theory of Ehresmann \ and Vandremeersch\cite{Ehresmann2}.
This way the states space \textit{X} can be seen as the subset $\left[
0,1\right]  ^{N}$ of \ $\mathbb{R}^{N}$ ($\mathbb{R}=$ set of real numbers, $N=$ dimension of \textit{X}), each point of
\textit{X} being a fuzzy category. Moreover we can define a smooth dynamical
system where the dynamics are given by adequate diffeomorphisms so that all
the concepts and techniques of the modern theory of dynamical systems can be
applied. Thus, it will be possible to determine several quantities, such as
the Lyapunov exponents, that will give us rich information about the systems
dynamical behavior\cite{Katok1}. It is worth noticing that the abstract theory
of dynamical systems in the context of categories has been already considered
by diverse authors but, within a pure theoretic spirit, as a tool to relate
different mathematical fields and concepts\cite{Behrisch1,Kontsevich1}.

In particular, in this work we focus into the application of the proposed
formalism to describe some aspects of the living single cell. The theoretical
study of the living single cell (specifically a minimal isolated bacterium)
has deserved increasing attention in the last forty years. The efforts have
mainly pointed out towards the development of models and mathematical methods
that facilitate the computer simulation of one such a synthetic
cell\cite{Zeigler1}. We can mention in this context the coarse-grained Cornell
\textit{Escherichia coli} B/r-A model\cite{Shuler1}. In models posed at a
coarse-grained level some finer details of interest can be incorporated in
specific modules, so we can talk of hybrid models\cite{Shuler2}.

The idea of modular biology\cite{Hartwell1} is what predominates in the route
towards the simulation of a whole-cell. The modules represent different types
of processes inside the cell and each one is handled with an accurate
mathematical tool. The integration of all the modules within a common
framework is then essential in order to simulate the whole-cell
behavior\cite{Covert1}. The results in this direction are very
promising\cite{Covert2,Covert3}.

In our fuzzy hierarchical categories approach we can contemplate different
degrees of detail: from finer molecular patterns to coarse-grained patterns to
patterns involving global cellular objects and functions. This is naturally
accounted by colimit binding processes. Actually our strategy is in some sense
the inverse one of that considered by bioinformaticians in their whole-cell
simulations. Whereas they, in a modular approach, resort as much as it is
possible to finer levels, we use the finer levels just to determine, by means
of colimits, objects and relations at higher levels and try to describe their
dynamics (reductionism vs. emergence\cite{Anderson1}).

A crucial point in the dynamical description of the living cell using our
fuzzy hierarchical categories is the assignation of the dynamics, say the
diffeomorphisms that transform with time categories into categories. In
principle the functional form of these maps, that account for the cell cycle,
is unknown for us. To solve this difficulty we propose a high-dimensional
generalization of the Ricker map\cite{Ricker1} with its parameters determined
by requiring that the resulting system dynamical behavior verifies fundamental
issues. In the first place of the requirements we consider the putative fact
that life develops at the edge of chaos\cite{Packard2, Kauffman1,Thurner1}, a
regime that, from the point of view of the theory of dynamical systems,
extends in between the ordered phase defined by period-2$^{n}$cycles and the
unpredictable chaos. It is known that this behavior manifests as that the
largest Lypunov exponent is zero. This should mean that the sensitivity to
initial conditions as well other system dynamical properties, such as the rate
of entropy increase, asymptotically follow a power law instead of an
exponential one\cite{Tsallis1,Latora1, Robledo1}. We additionally demand that
the composition law between the category fuzzy elements be verified.

The paper is organized as follows. In the next section we introduce the fuzzy
hierarchical categories. First we review the basic concepts of categories in
general and hierarchical categories in particular as considered by Ehresmann
\ and Vandremeersch\cite{Ehresmann1,Ehresmann2}. Then the main definitions and
properties of fuzzy sets as given by Zadeh\cite{Zadeh1} are shown for,
finally, put both, the hierarchical categories and the fuzzy sets, together in
order to define the fuzzy hierarchical categories, the main objects in our
description. Section III is devoted to remember the principal concepts and
definitions regarding the general theory of dynamical systems. The notion of
maximum Lyapunov exponent is reviewed since it will be an important tool in
our work. We also discuss the application, in general, of these ideas to
probability spaces whose elements are the fuzzy hierarchical categories that
we have introduced before. The\ dynamical theory for fuzzy hierarchical
categories developed in previous Sections is used in Section IV to describe,
in particular, the living single cell. We present a hierachical scheme for the
category that represent the state of a single cell at a given time as a
coarse-grained level pattern whose elements are colimits of patterns\ at a
finer level. These patterns represents the main modules in which we consider
the cell can be partitioned: mater and energy generation (metabolism),
information storage (DNA), information translation (RNA) and information
realization (proteins). In turn, the colimit of the coarse-grained pattern
gives at the higher hierarchical level the single functional cell. We consider
in our hierarchical model the possibility of binary fission. The diagram also
help us to propose the parametrized map\ that transforms the categories with
time and we discuss the parameters determination in order that the system
evolves at the on-set of the chaos. In this Section we also consider a
simplification of the original diagram that make easier the calculations. This
diagram can be taken as a toy model of the real single cell.

\section{Fuzzy hierarchical categories}

This Section is devoted to introduce the main objects in our description of
complex systems: the fuzzy hierarchical categories.

\subsection{Categories}

\begin{figure}[ht]
 \centering
 \includegraphics[width=\textwidth]{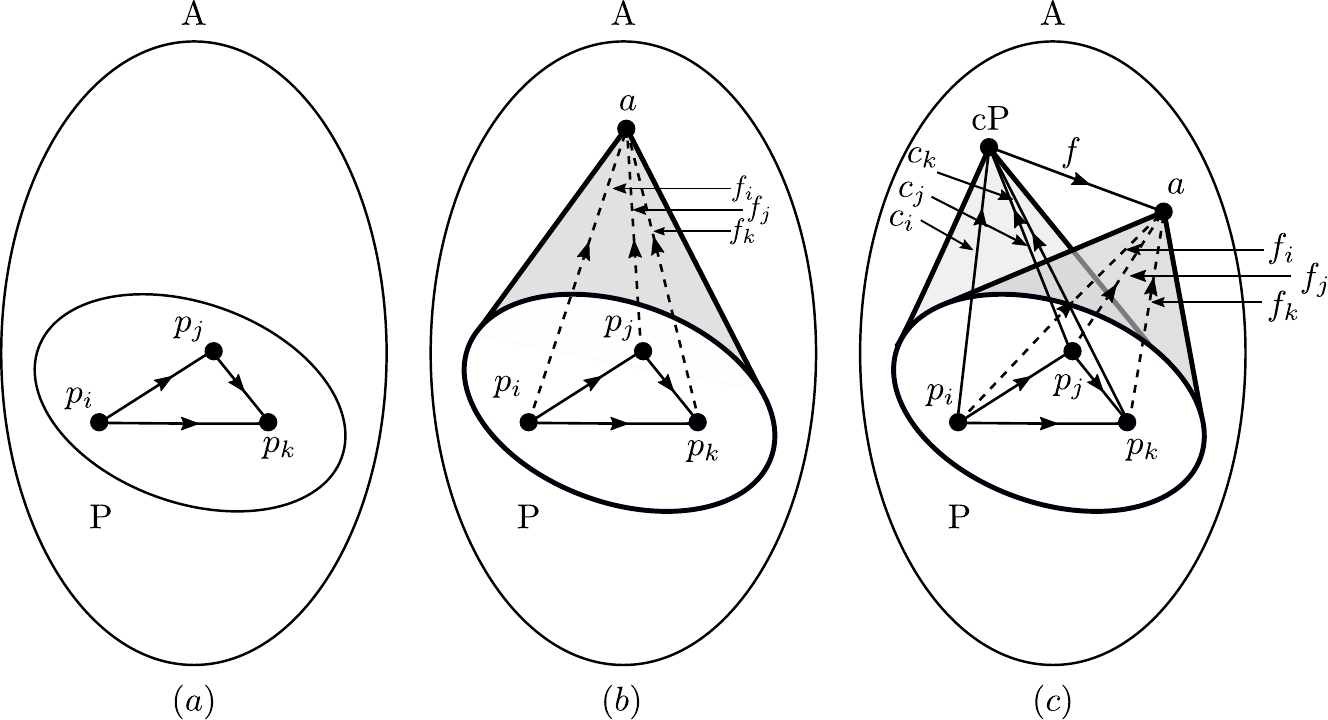}
 \caption{\textit{a}) Pattern P in the category A; \textit{b})
Collective link $\left(  f_{i}\right)  _{i\in\mathbf{I}}$ ($\mathbf{I=}%
\left\{  i,j,k\right\}  $) from the pattern P to the object $a$\textit{ }in
the category\textit{ }A; \textit{c}) Colimit cP of the pattern P in the
category A. The collective link from P to cP is $\left(  c_{i}\right)
_{i\in\mathbf{I}}$ ($\mathbf{I=}\left\{  i,j,k\right\}  $). We also show the
collective link $\left(  f_{i}\right)  _{i\in\mathbf{I}}$ ($\mathbf{I=}%
\left\{  i,j,k\right\}  $) from the pattern P to the object \textit{a }in the
category\textit{ }A that binds into the unique link $f:$cP$\rightarrow
$\textit{a }such that $f_{i}=f\cdot c_{i}$ for each $i\in\mathbf{I}$.}
 \label{fig:01}
\end{figure}

We start considering categories in general. Following very close the work by
Ehresmann \ and Vandremeersch\cite{Ehresmann2}, we give a few definitions to
introduce the subject.

\textbf{Definition 1: }A \textit{directed graph} $G$ is a set of objects
(\textit{vertices}) and a set of arrows (\textit{directed edges}) from a
vertex $a$ to a vertex $b$, denoted by $f:a\rightarrow b$. Here $a$ is the
\textit{source} of the arrow $f$ and $b$ its \textit{target}.

\textbf{Definition 2: }\ A \textit{category} $\mathbf{A}$ is a pair
constituted of a graph and an internal law of composition on the graph. The
law associates to successive arrows $f:a\rightarrow b$ and $g:b\rightarrow c$
a third ar$\operatorname{row}$ of the graph\ $h=g\cdot f:a\rightarrow c$ so
that the following properties are verified:

i) \textit{Associativity. }Given \ $f:a\rightarrow b$, $g:b\rightarrow c$
and\ $h:c\rightarrow d$, the composites $\left(  h\cdot g\right)  \cdot
f:a\rightarrow d$ and $h\cdot\left(  g\cdot f\right)  :$ $a\rightarrow d$ are
equal so we can unambiguously write $h\cdot g\cdot f:$ $a\rightarrow d$.

ii) \textit{Identity. }For\textit{ }each vertex\textit{ }$a$ in the graph
there is a closed arrow $i_{a}:a\rightarrow a$ called the \textit{identity of
}$a$. The identity is such that for any other arrow $f:a\rightarrow a$ is
$f\cdot i_{a}:a\rightarrow a=$ $i_{a}\cdot f:a\rightarrow a$\ $=f:a\rightarrow
a$.

\textbf{Definition 3: }A \textit{pattern }P in the category $\mathbf{A}$ (Fig.
1\textit{a}) is a family $\left(  p_{i}\right)  _{i\in\mathbf{I}}$ of objects
of $\mathbf{A}$ indexed by a finite set of indices $\mathbf{I}$. The objects
$p_{i}$ are the \textit{components} of the pattern and are such that, for each
each pair of indices $\left(  i\text{, }j\right)  $ there is a set of links
from $p_{i}$ to $p_{j}$ called the \textit{distinguished links}.

\textbf{Definition 4: }Let P be a pattern in the category $\mathbf{A}$. A
\textit{collective link }from P towards\textit{ }an object\textit{ }$a$ of
$\mathbf{A}$ (Fig.1\textit{b}) is a family $\left(  f_{i}\right)
_{i\in\mathbf{I}}$ of individual links of $\mathbf{A}$ such that associated to
each index $i$ of the pattern is a link $f_{i}$ from the component $p_{i}$ to
$a$.

\textbf{Definition 5: }An object of a category $\mathbf{A}$ is called the
\textit{colimit }of a pattern P in $\mathbf{A}$ (denoted cP) (see Fig.
1\textit{c}) if the following two conditions are satisfied:

i) there exists\textit{ }a collective link $\left(  c_{i}\right)
_{i\in\mathbf{I}}$ called the \textit{collective binding link }($c_{i}$ is the
\textit{binding link} from $p_{i}$ to cP );

ii) each collective link $\left(  f_{i}\right)  _{i\in\mathbf{I}}$ from the
pattern P to any object $a$ of $\mathbf{A}$ binds into one and only one link
$f$ from cP to $a$ which verifies the relations%

\[
f_{i}=f\cdot c_{i}\text{ for each index }i\in\mathbf{I}.
\]

\subsection{Hierarchical categories}

\begin{figure}[t!]
 \centering
 \includegraphics[height=0.7\textheight]{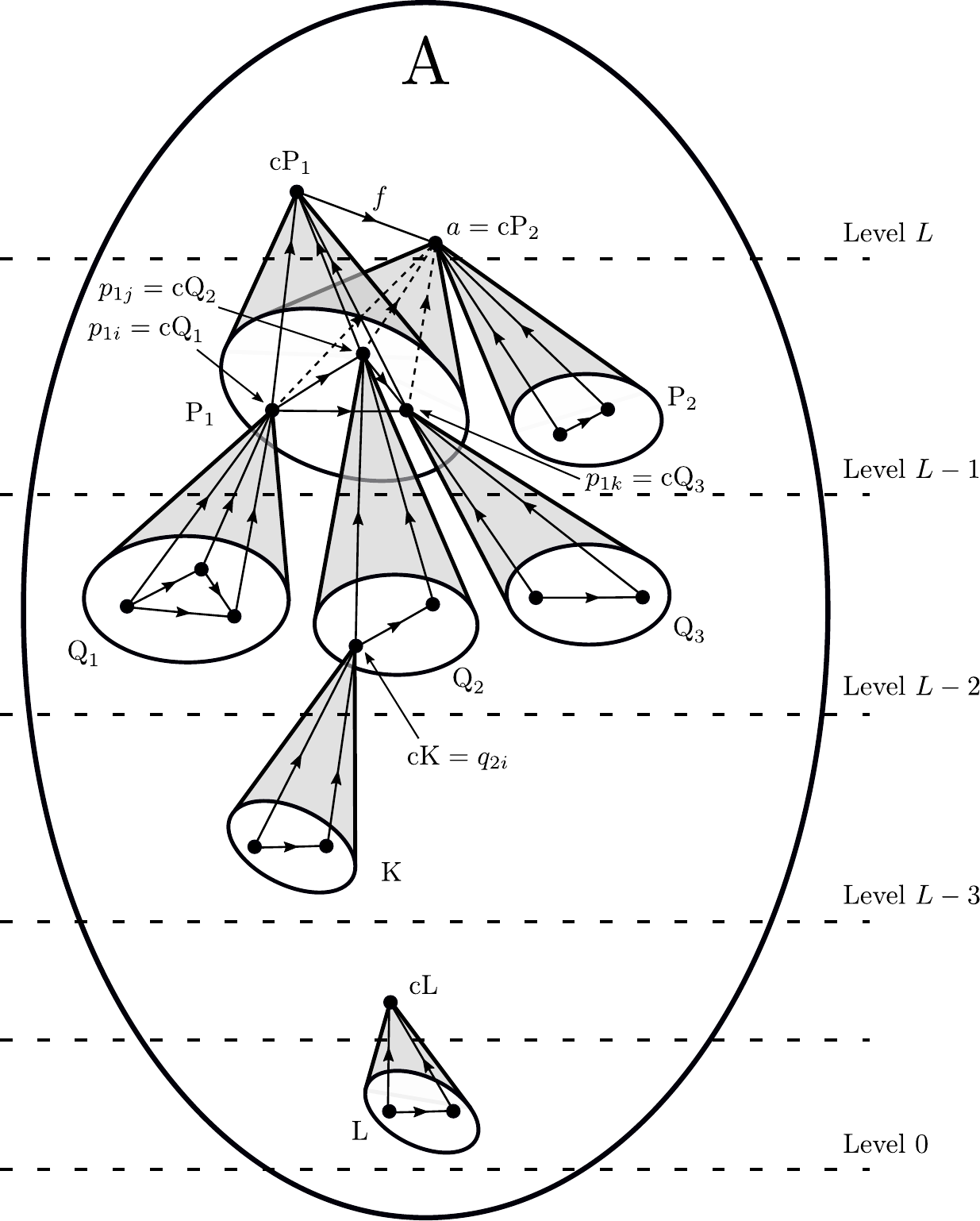}
 \caption{Hierarchical category.}
 \label{fig:02}
\end{figure}

\textbf{Definition 6: }A \textit{hierarchical category }is a category
$\mathbf{A}_{h}$ whose objects are partitioned into a finite sequence of
levels $0,1,\cdots,L$ , in such a way that any object $a$ of the level
$\ell+1$ is the colimit in $\mathbf{A}_{h}$ of at least one pattern P included
in levels lower than $\ell+1$ (Fig. 2).

The previous definitions guarantee that the relations among the components at
a given level are coherent with the relations among components at lower levels.

\subsection{Fuzzy sets}

\subsubsection{\textbf{Definition}}

We give a definition of fuzzy sets and some of their main
properties\cite{Zadeh1}.

\textbf{Definition 7: }Let $X$ be a\ collection of objects. We define a fuzzy
set $\tilde{A}$ as a set of pairs:%

\begin{equation}
\tilde{A}=\left\{  \left(  x,\mathbf{\chi}_{A}(x)\right)  \text{: }x\in
X\text{, }\mathbf{\chi}_{A}:X\rightarrow\left[  0,1\right]  \right\}  .
\tag{1}\label{1}%
\end{equation}

The function $\mathbf{\chi}_{A}:X\rightarrow\left[  0,1\right]  $ is the
membership (characteristic) function for the elements of $X$ in $A$.

\subsubsection{\textbf{Properties}}

a) Equality%

\begin{equation}
\left(  \tilde{A}=\tilde{B}\right)  \Longleftrightarrow\forall x:x\in
X\Longrightarrow\mathbf{\chi}_{A}(x)=\mathbf{\chi}_{B}(x). \tag{2}\label{2}%
\end{equation}

b) Containment (inclusion)%

\begin{equation}
\left(  \tilde{A}\subset\tilde{B}\right)  \Longleftrightarrow\forall x:x\in
X\Longrightarrow\mathbf{\chi}_{A}(x)\leq\mathbf{\chi}_{B}(x). \tag{3}\label{3}%
\end{equation}

c) Union%

\begin{equation}
\tilde{C}=\left(  \tilde{A}\cup\tilde{B}\right)  \Longleftrightarrow\forall
x:x\in X\Longrightarrow\mathbf{\chi}_{C}(x)=\max\left[  \mathbf{\chi}%
_{A}(x)\text{, }\mathbf{\chi}_{B}(x)\right]  . \tag{4}\label{4}%
\end{equation}

d) Intersection%

\begin{equation}
\tilde{C}=\left(  \tilde{A}\cap\tilde{B}\right)  \Longleftrightarrow\forall
x:x\in X\Longrightarrow\mathbf{\chi}_{C}(x)=\min\left[  \mathbf{\chi}%
_{A}(x)\text{, }\mathbf{\chi}_{B}(x)\right]  . \tag{5}\label{5}%
\end{equation}

e) Complement%

\begin{equation}
\tilde{A}_{c}=\left\{  \left(  x,\mathbf{\chi}_{A_{c}}(x)\right)  \text{:
}x\in X\text{, }\mathbf{\chi}_{A_{c}}(x)=1-\mathbf{\chi}_{A}(x)\right\}  .
\tag{6}\label{6}%
\end{equation}

f) Empty fuzzy set:%

\[
\tilde{\Phi}=\left\{  \left(  x,\mathbf{\chi}_{\Phi}(x)\right)  \text{: }x\in
X\text{ }\Longrightarrow\text{ }\mathbf{\chi}_{\Phi}(x)=0\right\}  .
\]

\textbf{Definition 8:} Two fuzzy sets are said disjoint if $\tilde{A}%
\cap\tilde{B}=\tilde{\Phi}$.

\subsection{\textbf{Fuzzy categories}}

We consider the case when each element $x$ of $X$ is an arrow $f$ together
with its source $a$ and its target $b$. We denote this composite object as
$a\overset{f}{\rightarrow}b$:\textbf{ }%
\begin{equation}
X:=\left\{  a\overset{f}{\rightarrow}b:\text{ }a,b\text{ are objects
(vertices), }f\text{ is an arrow (edge)}\right\}  . \tag{7}\label{7}%
\end{equation}

We assume that $\left\vert X\right\vert =N$ (large or even $\infty$) and define:

i) Equality of elements of $X$%

\begin{equation}
\left(  a\overset{f}{\rightarrow}b\right)  =\left(  a^{\prime}\overset
{f^{\text{ }\prime}}{\rightarrow}b^{\prime}\right)  \text{ iff }a=a^{\prime
}\text{, }b=b^{\prime}\text{ and }f=f^{\text{ }\prime}. \tag{8}\label{8}%
\end{equation}

ii) Composition (product) of elements of $X$: the product of $\left(
a\overset{f}{\rightarrow}b\right)  $ with $\left(  b\overset{g}{\rightarrow
}c\right)  $ gives%

\begin{equation}
\left(  a\overset{f}{\rightarrow}b\right)  \bullet\left(  b\overset
{g}{\rightarrow}c\right)  =\left(  a\overset{g\cdot f}{\rightarrow}c\right)  .
\tag{9}\label{9}%
\end{equation}

Within this context, we see that a category $\mathbf{A}$ can be considered as
a subset of elements of $X$ (Eq. \ref{7}) together with the composition law
(Eq. \ref{9}) for all the \textit{paths} of length $2$, say for all the
successive elements of the form $\left(  a\overset{f}{\rightarrow}b\text{,
}b\overset{g}{\rightarrow}c\right)  $.

\textbf{Definition 9: }A category $\mathbf{A}$ is a fuzzy category (denoted
$\mathbf{\tilde{A}}$) if:

i) each element $x=a\overset{f}{\rightarrow}b$ of $X$ (Eq. \ref{7}) is also an
element of $\mathbf{A}$ and has associated a membership value $\mathbf{\chi
}_{\mathbf{A}}\left(  x\right)  =\mathbf{\chi}_{\mathbf{A}}\left(
a\overset{f}{\rightarrow}b\right)  $.

ii) for each triple of elements of $X$ of the form $\left(  a\overset
{f}{\rightarrow}b\text{, }b\overset{g}{\rightarrow}c\text{, }a\overset{g\cdot
f}{\rightarrow}c\right)  $ is%

\begin{equation}
\mathbf{\chi}_{\mathbf{A}}\left(  a\overset{g\cdot f}{\rightarrow}c\right)
=\min\left[  \mathbf{\chi}_{\mathbf{A}}\left(  a\overset{f}{\rightarrow
}b\right)  \text{, }\mathbf{\chi}_{\mathbf{A}}\left(  b\overset{g}%
{\rightarrow}c\right)  \right]  . \tag{10}\label{10}%
\end{equation}

We denote with $\mathbf{\tilde{C}}$ the set of all the fuzzy categories
generated from $X$ as defined in Eq.(\ref{7}). We can think $\mathbf{\tilde
{A}}$ as a $N$-dimensional point ($\mathbf{\tilde{A}\in}\left[  0,1\right]
^{N}\subset\mathbb{R}^{N}$ ) in the phase space $\mathbf{\tilde{C}}$:

\begin{equation}
\mathbf{\tilde{A}=}\text{ }\left(  \mathbf{\chi}_{\mathbf{A}}(x_{1}),\text{
}\mathbf{\chi}_{\mathbf{A}}(x_{2}),\cdots,\mathbf{\chi}_{\mathbf{A}}
(x_{N})\text{ }\right)  \tag{11}\label{11}
\end{equation}

\subsection{Fuzzy hierarchical categories}

A fuzzy hierarchical category $\mathbf{\tilde{A}}_{h}$ is a hierarchical
category $\mathbf{A}_{h}$ that, in addition, is fuzzy. We denote with
$\mathbf{\tilde{C}}_{h}$ the set of all the fuzzy hierarchical categories.

A fuzzy hierarchical category of levels $0,1,\cdots,L$ can be written as the
union of fuzzy sets (graphs) $\tilde{A}_{\ell\ell^{\prime}}$($\ell
,\ell^{\prime}=0,1,\cdots,L$) which are disjoint by pairs:%

\begin{equation}
\mathbf{\tilde{A}}_{h}=\bigcup\limits_{\ell,\ell^{\prime}}\tilde{A}_{\ell
\ell^{\prime}}, \tag{12}\label{12}%
\end{equation}
with $\tilde{A}_{\ell\ell^{\prime}}\cap\tilde{A}_{kk^{\prime}}=\tilde{\Phi}$
for $(\ell,\ell^{\prime})$ $\neq(k,k^{\prime})$. The elements of $\tilde
{A}_{\ell\ell^{\prime}}$ with membership $\mathbf{\chi}_{\tilde{A}_{\ell
\ell^{\prime}}}(x)\neq0$ are arrows that have their source at the level $\ell$
and their target at the level $\ell^{\prime}$. In particular the elements of
$\tilde{A}_{\ell\ell}$ are arrows whose source and target are both contained
in the same level $\ell$. It is worth mentioning that, in general, the graphs
$\tilde{A}_{\ell\ell^{\prime}}$ are not categories because they could not
verify internally the composition law.

\section{\bigskip Dynamical systems in fuzzy categories}

\subsection{\textbf{Measure preserving dynamical system}}

First we remember the concept of measure preserving dynamical systems in
general\cite{Katok1}. It is a quadruple $\left(  \text{X},\mathcal{B}%
,\mu,\mathcal{T}\right)  $ consisting of

i) a probability space $\left(  \text{X},\mathcal{B},\mu\right)  $ where $\mu$
is a measure and $\mathcal{B}$ is the Borel $\sigma$-algebra on X.

ii) a $\mathcal{B}$-measurable\ $G$-action $\mathcal{T}=$ $\left(  T^{g}:g\in
G\right)  $\ so that \ $T^{g}\mu=\mu$.\ \ \ \ \ \ \ \ \ \ \ \ \ \ \ \ \ \ \ \ \ \ \ \ \ \ \ \ \ \ \ \ \ \ \ \ \ \ \ \ \ \ \ \ \ \ \ \ \ 

We apply this definition to the categories we have introduced. Thus, we have
the quadruple $\left(  \mathbf{\tilde{C}},\mathcal{B},\mu,\mathcal{T}\right)
$ and consider $G=\mathbb{N}$ and monoid actions such that $T:\mathbf{\tilde{C}}\rightarrow\mathbf{\tilde
{C}}$ assigns to the fuzzy category $\mathbf{\tilde{A}\in}\left[  0,1\right]
^{N}$ another fuzzy category $\mathbf{\tilde{B}\in}\left[  0,1\right]  ^{N}$
(a map that transforms categories into categories is usually called a
functor):
\begin{equation}
\mathbf{\tilde{B}=}T(\mathbf{\tilde{A}}).\tag{13}\label{13}%
\end{equation}
The functor $T$ transforms the pairs $\left(  x_{i},\mathbf{\chi}_{A}%
(x_{i})\right)  \in$ $\mathbf{\tilde{A}}$ into the pairs $\left(
x_{i},\mathbf{\chi}_{B}(x_{i})\right)  \in\mathbf{\tilde{B}}$ with the new
membership values depending of the old ones, say%

\begin{equation}
\mathbf{\chi}_{\mathbf{B}}(x_{i})=T_{i}\left(  \mathbf{\chi}_{\mathbf{A}%
}(x_{i});\text{ }\left\{  \mathbf{\chi}_{\mathbf{A}}(x_{j})\right\}  _{j\neq
i}\text{ }\right)  \text{ \ \ }i=1,2,\cdots,N, \tag{14}\label{14}%
\end{equation}
in such a way that the composition law is conserved: if\ $\ \left(
a\overset{g\cdot f}{\rightarrow}c\right)  =\left(  a\overset{f}{\rightarrow
}b\right)  \bullet\left(  b\overset{g}{\rightarrow}c\right)  $ so%

\begin{equation}
\mathbf{\chi}_{\mathbf{A}}\left(  a\overset{g\cdot f}{\rightarrow}c\right)
=\min\left[  \mathbf{\chi}_{\mathbf{A}}\left(  a\overset{f}{\rightarrow
}b\right)  \text{, }\mathbf{\chi}_{\mathbf{A}}\left(  b\overset{g}%
{\rightarrow}c\right)  \right]  , \tag{15}\label{15}%
\end{equation}
then also is%

\begin{equation}
\mathbf{\chi}_{\mathbf{B}}\left(  a\overset{g\cdot f}{\rightarrow}c\right)
=\min\left[  \mathbf{\chi}_{\mathbf{B}}\left(  a\overset{f}{\rightarrow
}b\right)  \text{, }\mathbf{\chi}_{\mathbf{B}}\left(  b\overset{g}%
{\rightarrow}c\right)  \right]  , \tag{16}\label{16}%
\end{equation}
or, symbolically, $T\left(  x_{i}\bullet x_{j}\right)  =T\left(  x_{i})\bullet
T(x_{j}\right)  $.

\subsection{\textbf{Lyapunov characteristic numbers}}

The Lyapunov exponents are a measure of the sensitivity of the system dynamics
to initial conditions\cite{Robinson1}. In one dimensional systems there is
just one and it can be expressed as the growth rate of the derivative of the
transformation. In our case we must consider differentiation in higher
dimensions. Also we must take into account that around a given point
$\mathbf{\tilde{B}}$ there are directions where the map can be an expansion
and others where it is a contraction.

We start introducing differentiation in higher dimensions. We assume that the
map $T$ is continuously differentiable: $T\in$ $C^{1}$. The partial
derivatives at a phase space point $\mathbf{\tilde{B}}$ can be condensed into
a single matrix $DT_{\mathbf{\tilde{B}}}$ of the form%

\begin{equation}
DT_{\mathbf{\tilde{B}}}=\left[  \frac{\partial T_{i}(\mathbf{\tilde{B}}%
)}{\partial\mathbf{\chi}_{\mathbf{A}}(x_{j})}\right]  \text{ \ \ \ }\left(
i,j=1,2,\cdots,N\right)  \tag{17}\label{17}%
\end{equation}
and%

\begin{equation}
DT_{\mathbf{\tilde{B}}}^{k}=\left[  \frac{\partial T_{i}^{k}(\mathbf{\tilde
{B}})}{\partial\mathbf{\chi}_{\mathbf{A}}(x_{j})}\right]  \text{
\ \ \ }\left(  i,j=1,2,\cdots,N\right)  \tag{18}\label{18}%
\end{equation}
for the $k$th-iteration.

The directions of expansion and contraction at a point $\mathbf{\tilde{B}}$
can be taken as infinitesimal displacements of $\mathbf{\tilde{B}}$ known as
tangent vectors. We can think of a tangent vector at $\mathbf{\tilde{B}}$ as
the derivative of a curve through $\mathbf{\tilde{B}}$. Thus if $\gamma
:\left(  -\delta\text{, }\delta\right)  $ $\rightarrow\mathbf{\tilde{C}}$ is a
differentiable curve with $\gamma\left(  0\right)  =$ $\mathbf{\tilde{B}}$,
then $\mathbf{v=}$ $\gamma^{\prime}\left(  0\right)  $ is a tangent vector at
$\mathbf{\tilde{B}}$, usually denoted $\mathbf{v}_{\mathbf{\tilde{B}}}$. We
call the set of all possible tangent vectors at $\mathbf{\tilde{B}}$ as the
tangent space at $\mathbf{\tilde{B}}$ and write\textsl{ }$\mathfrak{T}%
_{\mathbf{\tilde{B}}}\mathbf{\tilde{C}}$. To determine lengths of the vectors
we consider the inner product on the tangent space $\left\langle \bullet\text{
},\bullet\right\rangle _{\mathbf{\tilde{B}}}:\mathfrak{T}_{\mathbf{\tilde{B}}%
}\mathbf{\tilde{C}\times}\mathfrak{T}_{\mathbf{\tilde{B}}}\mathbf{\tilde
{C}\rightarrow\mathbb{R}}$ and define the Riemannian norm $\left\Vert \bullet\right\Vert
_{\mathbf{\tilde{B}}}=\left\langle \bullet\text{ },\bullet\right\rangle
_{\mathbf{\tilde{B}}}^{1/2}$. 

Using the tangent vectors we can define directional derivatives
$DT_{\mathbf{\tilde{B}}}\mathbf{v}$ whose $i$th-coordinate is given by%

\begin{equation}
\left(  DT_{\mathbf{\tilde{B}}}\mathbf{v}\right)  _{i}=\sum\limits_{j=1}%
^{N}\left[  \frac{\partial T_{i}(\mathbf{\tilde{B}})}{\partial\mathbf{\chi
}_{\mathbf{A}}(x_{j})}\right]  \mathbf{v}_{j}\text{ \ \ \ }\left(
i=1,2,\cdots,N\right)  \tag{19}\label{19}%
\end{equation}

We already have the basic elements needed to define the Lyapunov
characteristic numbers:

\textbf{Definition 10: }Let $T:\mathbf{\tilde{C}}\rightarrow\mathbf{\tilde{C}%
}$ (assumed as a diffeomorphism on $\mathbf{\tilde{C}}$) be the system
dynamics and let $\left\Vert \bullet\right\Vert $ be the Riemannian norm on
tangent vectors. For each $\mathbf{\tilde{B}\in}$ $\mathbf{\tilde{C}}$ and
$\mathbf{v\in}$ $\mathfrak{T}_{\mathbf{\tilde{B}}}\mathbf{\tilde{C}}$, let%

\begin{equation}
\lambda\left(  \mathbf{\tilde{B}}\text{, }\mathbf{v}\right)  =\lim
_{k\rightarrow\infty}\frac{1}{k}\log\left\Vert DT_{\mathbf{\tilde{B}}}%
^{k}\mathbf{v}\right\Vert \tag{20}\label{20}%
\end{equation}
whenever the limit exists.

For almost all states $\mathbf{\tilde{B}}$, the Multiplicative Ergodic Theorem
of Oseledets\cite{Oseledets1} says that this limit exists for all
$\mathbf{v\in}$ $\mathfrak{T}_{\mathbf{\tilde{B}}}\mathbf{\tilde{C}}$. Also it
is shown that there exists a basis $\left(  \mathbf{v}^{1},\mathbf{v}%
^{2},\cdots,\mathbf{v}^{s(\mathbf{\tilde{B}})}\right)  $ of $\mathfrak{T}%
_{\mathbf{\tilde{B}}}\mathbf{\tilde{C}}$ such that%

\begin{equation}
\sum\limits_{\alpha=1}^{s(\mathbf{\tilde{B}})}\lambda\left(  \mathbf{\tilde
{B}}\text{, }\mathbf{v}^{\alpha}\right)  =\inf_{\Lambda}\sum\limits_{\alpha
=1}^{s(\mathbf{\tilde{B}})}\lambda\left(  \mathbf{\tilde{B}}\text{,
}\mathbf{\tilde{v}}^{\alpha}\right)  \tag{21}\label{21}%
\end{equation}
where%
\[
\Lambda=\left\{  \left(  \mathbf{\tilde{v}}^{1},\mathbf{\tilde{v}}^{2}%
,\cdots,\mathbf{\tilde{v}}^{s(\mathbf{\tilde{B}})}\right)  :\left(
\mathbf{\tilde{v}}^{1},\mathbf{\tilde{v}}^{2},\cdots,\mathbf{\tilde{v}%
}^{s(\mathbf{\tilde{B}})}\right)  \text{ is a basis of }\mathfrak{T}%
_{\mathbf{\tilde{B}}}\mathbf{\tilde{C}}\right\}
\]
As $\mathbf{v}$ varies in $\mathfrak{T}_{\mathbf{\tilde{B}}}\mathbf{\tilde{C}%
}$, $\lambda\left(  \mathbf{\tilde{B}}\text{, }\mathbf{v}\right)  $ takes only
values of the set $\left\{  \lambda\left(  \mathbf{\tilde{B}}\text{,
}\mathbf{v}^{\alpha}\right)  \right\}  _{1\leq\alpha\leq s(\mathbf{\tilde{B}%
})}$. The number $\lambda\left(  \mathbf{\tilde{B}}\text{, }\mathbf{v}\right)
$ is called the Lyapunov characteristic number of the vector $\mathbf{v\in}$
$\mathfrak{T}_{\mathbf{\tilde{B}}}\mathbf{\tilde{C}}$ and the numbers
$\lambda\left(  \mathbf{\tilde{B}}\text{, }\mathbf{v}^{\alpha}\right)  $,
which depend only on the map $T$ and the point $\mathbf{\tilde{B}}$, are
called the Lyapunov characteristic numbers of the map $T$ at $\mathbf{\tilde
{B}}$.

We denote $\lambda\left(  \mathbf{\tilde{B}}\text{, }\mathbf{v}^{\alpha
}\right)  =$ $\lambda_{\alpha}\left(  \mathbf{\tilde{B}}\right)  $,
$1\leq\alpha\leq s(\mathbf{\tilde{B}})$ and assume%

\begin{equation}
\lambda_{1}\left(  \mathbf{\tilde{B}}\right)  >\lambda_{2}\left(
\mathbf{\tilde{B}}\right)  >\cdots>\lambda_{s(\mathbf{\tilde{B}})}\left(
\mathbf{\tilde{B}}\right)  \text{.} \tag{22}\label{22}%
\end{equation}
The largest $\lambda_{\alpha}\left(  \mathbf{\tilde{B}}\right)  $ will be
called $\lambda_{\max}\left(  \mathbf{\tilde{B}}\right)  $: $\lambda
_{1}\left(  \mathbf{\tilde{B}}\right)  =$ $\lambda_{\max}\left(
\mathbf{\tilde{B}}\right)  $.

\section{Living single cell as a dynamical system in fuzzy hierarchical
categories}

\subsection{General model}

\begin{figure}[t!]
 \centering
 \includegraphics[width=\textwidth]{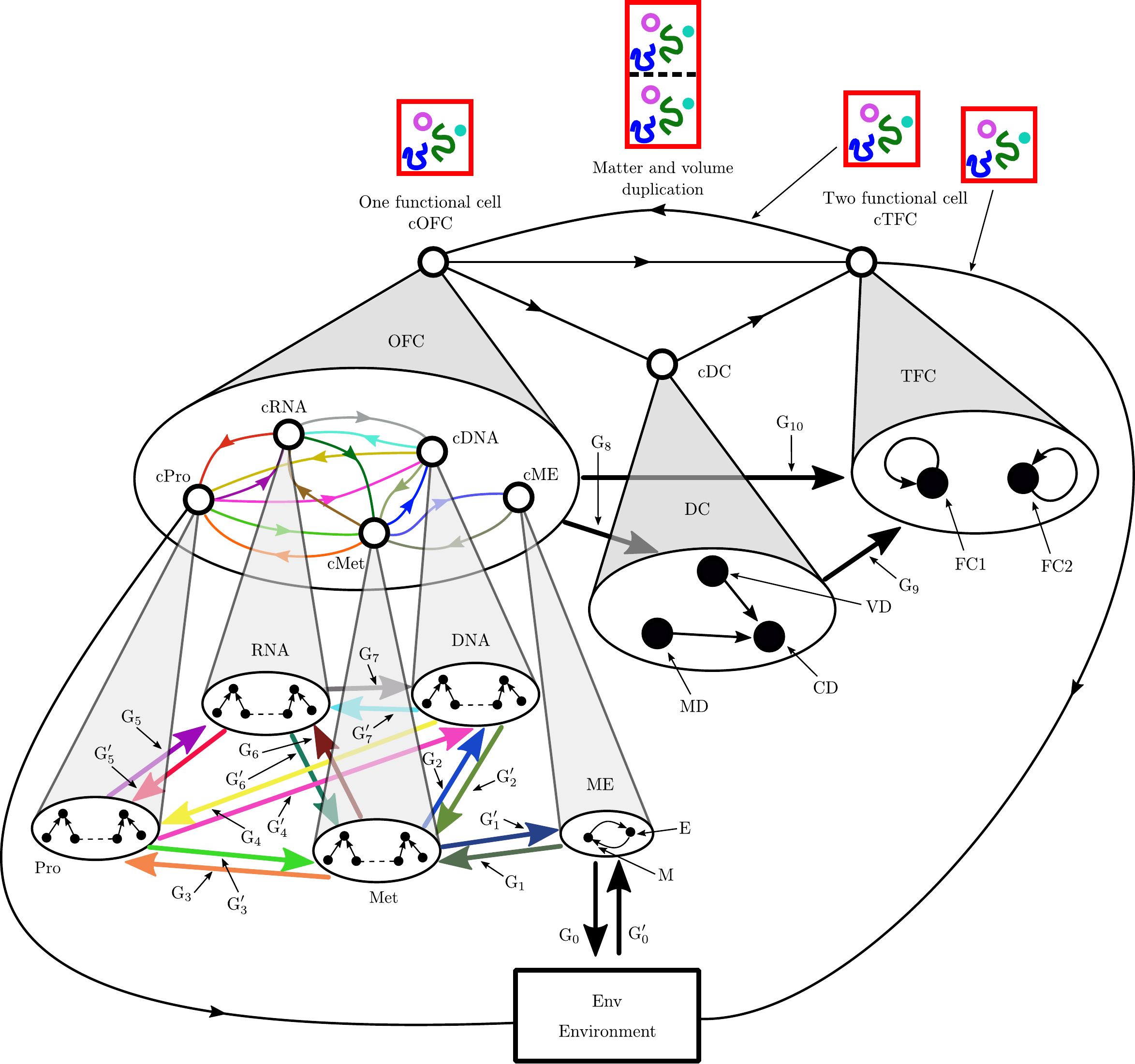}
 \caption{A hierarchical category to represent the living single cell
(see text for explanation).}
 \label{fig:03}
\end{figure}

Most of the attempts done to describe single cells and bacteria from a
theoretical and computational point of view\cite{Covert1,Covert2,Covert3},
apply the modular cell biology approach\cite{Hartwell1}. In general, the
description is at coarse-grained level and each module is treated with more
detail (finer-level) using appropriate mathematical tools. The modules are
finally integrated into a common computational frame\cite{Covert1}.

In our description of the living single cell here we use instead the dynamical
system theory for fuzzy hierarchical categories developed in previous
Sections. The states of the cell are represented by fuzzy categories
$\mathbf{\tilde{A}\in\tilde{C}=}\left[  0,1\right]  ^{N}$ and we consider a
dynamical system $\left(  \mathbf{\tilde{C}},\mathcal{B},\mu,\mathcal{T}%
\right)  $ with the action $\mathcal{T}=$ $\left(  T^{n}:n\in\mathbb{N}\right)  $. All the states $\mathbf{\tilde{A}}$ are supported by $X$ (Eq.
\ref{7}) say, all they have the same elements $x=\left(  a\overset
{f}{\rightarrow}b\right)  \mathbf{\in}X$ and differ among them in the
membership degree of these elements (We call the vertices $a$ and $b$\textit{
}the\textit{ components} and the composite objects $\left(  a\overset
{f}{\rightarrow}b\right)  $ the \textit{elements} of $\mathbf{\tilde{A}}$ or
of their patterns).

In Fig. 3 we present the scheme we propose for the single cell which is valid
for all the states (categories of $\mathbf{\tilde{C}}$) because, as was
already mentioned, the different states distinguish themselves by just the
membership degree of their elements. We can recognize three main levels.

The lower level (molecular level) shows five patterns labeled ME (for Matter
and Energy); Met (for metabolism); DNA; RNA and Pro (for proteins). Except for
ME in the other four patterns we have schematically drawn distinguished links
among their components that, just for illustration, we have pictured with the
same form. As we will see, the fine details at molecular and atomic level
actually will be irrelevant for our analysis here. The pattern ME receives
special attention because is through it that the cell relates with the
environment, a relation that we want to emphasize. With DNA we mean all the
information-carrying genetic apparatus, whereas RNA involves the whole
translation machinery that transforms the genetic information into the
complete set of ubiquitous polypeptides and\ proteins. On the other hand Met
refers to the complex network of chemical reactions that get the molecules and
energy necessary for the diverse modules can work.

The intermediate level (coarse-grained level)\ has three patterns denoted OFC
(one functional cell); DC (duplicated cell) and TFC (two functional cells).
The components of the pattern OFC are the colimits of the patterns at the
lower level. We indicate in this figure the colimits with a small open circle.
In all the cases the collective binding link from each pattern component to
the corresponding colimit has been omitted in order to simplify the drawing.

The third level (cellular level) has only one pattern that consists of three
components (cOFC, cDC, cTFC) which are just the colimits of the three patterns
of the second level.

The patterns of a given level are related among them by mean of clusters whose
definition we remember now\cite{Ehresmann2}:

\begin{figure}[t!]
 \centering
 \includegraphics[width=\textwidth]{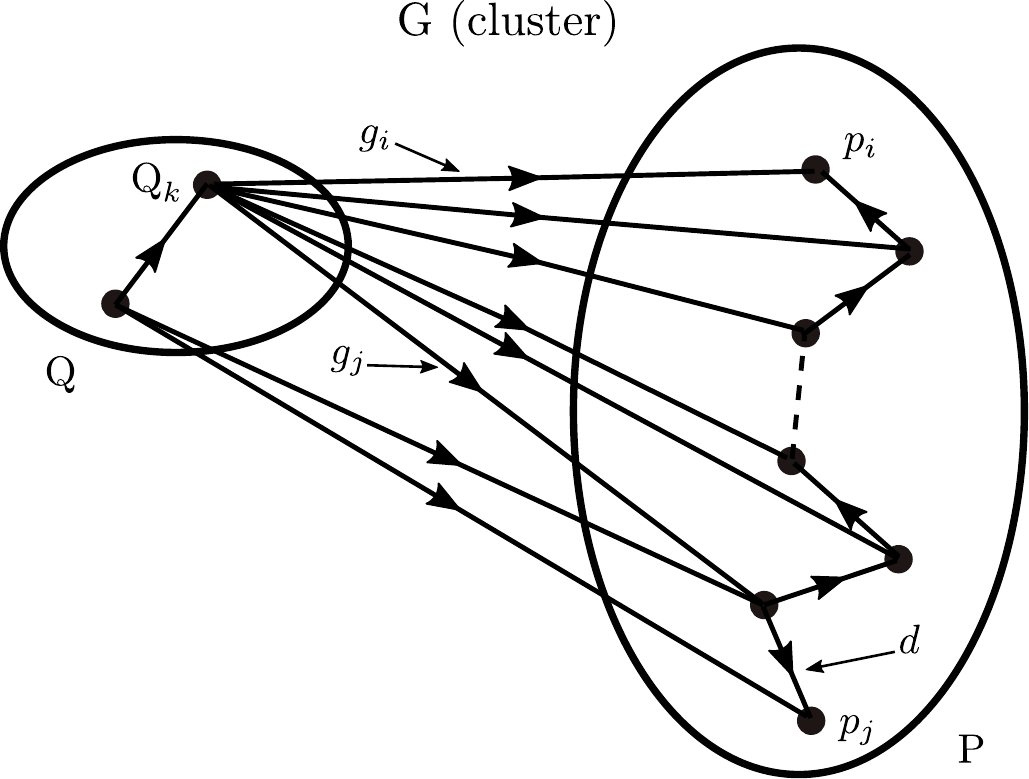}
 \caption{A cluster from the pattern Q to the pattern P.}
 \label{fig:04}
\end{figure}

\textbf{Definition 11}(see Fig.4)\textbf{. }Given two\textbf{ }patterns P and
Q in a category, a cluster from Q to P is a maximal set of links between
components of these patterns that satisfies the following conditions:

i) For each index $k$, the component Q$_{k}$ of Q has at least one link to a
component of P and \ if there are several such links, they are correlated by a
zigzag of distinguished links of P.

ii) The composite of a link of the cluster with a distinguished link of P (as
$d$ in Fig. 4), or of a distinguished link of Q (v.g. the only one shown in
Fig. 4) with a link of the cluster, also belongs to the cluster.

For simplicity, in Fig. 3, we have represented the whole set of links in each
cluster with just a broad arrow labeled \textbf{G}$_{\text{i}}$ and
\textbf{G}$_{\text{i}}^{\prime}$ (i $=0,1,2,\cdots,7$) and \textbf{G}$_{8}%
,$\textbf{G}$_{9}$ and \textbf{G}$_{10}$. Note that in correspondence with the
clusters in a lower level we have links between the colimits of the
corresponding patterns. In particular, in Fig. 3, the clusters between the
pairs of patterns ME, Met, DNA, RNA and Pro have been drawn with different
colors which are the same ones used for the links between colimits in the
upper level.

In the APPENDIX A we explicit the elements considered in Fig. 3 for our model
of living cell. The description is limited for simplicity to the intermediate
and higher levels so that the \textit{collective binding links \ }$\left(
c_{i}\right)  _{i\in\mathbf{I}}$ \ from the patterns Met, DNA, RNA and Pro in
the lower level to the corresponding colimits cMet, cDNA, cRNA and cPro,
respectively, are not taken into account. In the same spirit we also ignore
the internal structure of the clusters in the lower level. This means that we
restrict ourselves to a coarse-grained description. This is a simplified
version of the more realistic model we could hypothetically construct
following the present approach (v.g. by introducing in detail the molecular
level). Below we consider an even simpler version (toy model) that should make
explicit calculations more accessible (Fig. 5).

Although the \textit{collective binding links \ }$\left(  c_{i}\right)
_{i\in\mathbf{I}}$ \ from the patterns OFC, CDF and TFC in the coarse-grained
level to the corresponding colimits cOFC, cDC and cTFC, respectively, are not
drawn in Fig. 3 they are explicitly considered in tables \ref{tbl:2a} - \ref{tbl:2f} of the APPENDIX
A. Besides in tables \ref{tbl:2a} - \ref{tbl:2f} we take into account the possibility that between a
source $a$ and a target $b$ be more than one arrow (which are differentiated
with prime symbols). These elements mean either the indirect interaction
between $a$ and $b$ via a third element $c$ or simply their direct interaction.

To give an explicit form for the map in Eq.\ref{14} that acts as the dynamic
in our representation of the living single cell as a dynamical system, we
previously define the subsets $\mathit{K}_{i}^{\dag}$ and $\mathit{K}%
_{i}^{\ddag}$ formed by those elements of the category that represents the
cell which, in addition, are neighbors of the element $x_{i}$. We consider
that a map $x_{r}$ is neighbor of the map $x_{i}$ if they have at least one
vertex in common or, in other words, we take as neighbors of an element
$\left(  a\overset{f}{\rightarrow}b\right)  $ all the elements which have at
least one of $a$ and $b$ as source or target. In particular, we consider that
each element is neighbor of itself. The subsets $\mathit{K}_{i}^{\dag}$ and
$\mathit{K}_{i}^{\ddag}$ differ in that, whereas each element $x_{k}$ of
$\mathit{K}_{i}^{\ddag}$ can be expressed as product of two other elements
$x_{k_{1}}$and $x_{k_{2}}$ of the category: $x_{k}=x_{k_{1}}\bullet x_{k_{2}}%
$, \ the elements of $\mathit{K}_{i}^{\dag}$ can not. Thus we take for the map:

\begin{align}
\mathbf{\chi}_{\mathbf{B}}(x_{i})  &  =T_{i}\left(  \mathbf{\chi}_{\mathbf{A}%
}(x_{i});\text{ }\left\{  \mathbf{\chi}_{\mathbf{A}}(x_{j})\right\}  _{j\neq
i}\text{ }\right)  \text{ \ }\nonumber\\
&  =\left\{
\begin{array}
[c]{c}%
\text{ }\left(  \sum\limits_{j\in\left\{  j:\text{ }x_{j}\in\mathit{K}%
_{i}^{\dag}\cup\mathit{K}_{i}^{\ddag}\right\}  }\varepsilon_{ij}\right)
^{-1}\left(  \text{\ }\sum\limits_{x_{j}\in\mathit{K}_{i}^{\dag}}%
\varepsilon_{ij}r_{j}\mathbf{\chi}_{\mathbf{A}}(x_{j})\exp\left\{
1-r_{j}\mathbf{\chi}_{\mathbf{A}}(x_{j})\right\}  \right.  \text{
\ \ \ \ \ \ \ \ \ \ \ \ \ \ \ \ \ \ \ \ \ \ \ \ \ \ \ \ \ \ \ \ \ \ \ \ \ \ \ \ \ \ \ \ \ \ \ \ \ }%
\\
\left.  +\sum\limits_{x_{k}\in\mathit{K}_{i}^{\ddag}}\varepsilon_{ik}r_{k}%
\min\left[  \mathbf{\chi}_{\mathbf{A}}(x_{k_{1}}),\mathbf{\chi}_{\mathbf{A}%
}(x_{k_{2}})\right]  \exp\left\{  1-r_{k}\min\left[  \mathbf{\chi}%
_{\mathbf{A}}(x_{k_{1}}),\mathbf{\chi}_{\mathbf{A}}(x_{k_{2}})\right]
\right\}  \right)  \text{
\ \ \ \ \ \ \ \ \ \ \ \ \ \ \ \ \ \ \ \ \ \ \ \ \ \ }\\
\text{
\ \ \ \ \ \ \ \ \ \ \ \ \ \ \ \ \ \ \ \ \ \ \ \ \ \ \ \ \ \ \ \ \ \ \ \ \ \ if
}x_{i}\text{ is not product of other two elements
\ \ \ \ \ \ \ \ \ \ \ \ \ \ \ \ \ \ \ \ \ \ \ \ \ \ \ \ \ \ \ }\\
\min\left[  \mathbf{\chi}_{\mathbf{B}}(x_{i_{1}}),\mathbf{\chi}_{\mathbf{B}%
}(x_{i_{2}})\right]  \text{
\ \ \ \ \ \ \ \ \ \ \ \ \ \ \ \ \ \ \ \ \ \ \ \ \ \ \ \ \ \ \ \ \ \ \ \ \ \ \ \ \ \ \ \ \ \ \ \ \ \ \ \ \ \ \ \ \ \ \ \ \ \ \ \ \ \ \ \ \ \ \ \ \ \ \ \ \ \ \ \ \ \ \ \ \ \ \ \ \ \ \ \ \ \ \ \ \ \ \ \ \ \ \ }%
\\
\text{ \ \ \ \ \ \ \ \ \ \ \ \ \ \ \ \ \ \ \ \ if }x_{i}=x_{i_{1}}\bullet
x_{i_{2}}\text{
\ \ \ \ \ \ \ \ \ \ \ \ \ \ \ \ \ \ \ \ \ \ \ \ \ \ \ \ \ \ \ \ \ \ \ \ \ \ \ \ \ \ \ \ \ \ \ \ }%
\end{array}
\right.  \tag{23}\label{23}%
\end{align}
where $i=1,2,\cdots,N$ and $\varepsilon_{ij}$ denotes the coupling between the
element $x_{i}$ and its neighbor $x_{j}$. We see that Eq. \ref{23} maps the
domain $\left[  0,1\right]  ^{N}$ into itself.

It is evident the resemblance of the map given by Eq. \ref{23} with the Ricker
map\cite{Ricker1}. In fact, Eq. \ref{23} can be thought of as a $N$%
-dimensional version of the one-dimensional Ricker map that additionally
includes the composition law of fuzzy categories. We recall that the Ricker
map (normalized in order to tranform into itself the interval $\left[
0,1\right]  $), say $x_{n+1}=rx_{n}\exp\left(  1-rx_{n}\right)  $ (see
APPENDIX B) is in turn, a generalized version of the well known
logistic-map\cite{Strogatz1} for population dynamics $x_{n+1}=ax_{n}\left(
1-x_{n}\right)  $. In both models, the population, initially small, has
abundant resources to its disposal and grows. However, since the resources are
limited, for a larger population the competition for food among its
individuals causes that the death rate becomes higher than the birth rate and
population decreases.

In our case the elements $x_{i}$ are the diverse activities and processes of
the cell and the membership degree $\mathbf{\chi}_{\mathbf{A}}(x_{i})$ can be
taken as a measure of their strength. For a given $x_{i}$, \ when the
membership values of itself and of its neighbors are small and increase we
expect that they favour the process $x_{i}$ and then $\mathbf{\chi
}_{\mathbf{B}}(x_{i})$ also grows. But, if these processes strengths follow
growing beyond those values that allow a synchronized workings, then they
become more a disturbance for $x_{i}$ than a positive contribution. A rough
analogy of this situation can be found in the transit of a city crowded
street. Each car individually can run at over say 150 Km/hour. However, given
a car in the crowded street it is influenced by the others (particulary its
neighbors) so will increase its velocity until an optimal one, say about 60
Km/hour, at which the car will move in harmony with the whole transit. If the
chosen car or someone of its neighborhood increase the velocity over the
optimal one then the considered car in particular and the transit in general
will be perturbed in some way (vg. crashes) and, as a result, their velocity
will diminish.

Our strategy consists, as we will discuss in next Subsection, into adjust the
parameters in such a way that relevant aspects of the model (as describing a
living cell) be fulfilled. However the huge number of parameters make any
attempt in this direction a very hard task. To take an idea of the problem we
remember that, in one, two and three dimensions, the logistic, H\'{e}non and
Lorenz classical dynamic maps have only one, two and three parameters,
respectively. Dynamics in high dimensions with larger number of parameters are
considered, for example, in ref. \cite{Albers1}

Any way, we consider that the model shown in Fig. 3 is still useful in the
sense of providing a graphic description of a very complex system which,
furthermore, can suggest simplified versions of the whole model (toy models)
that allow to numerically study the single cell focusing into particular
aspects of its behavior. In this spirit, in Fig. 5, we present a toy model of
the living cell that emphasizes the binary fission. For this model, whose
state space has dimension $N=4$, the map in Eq. \ref{23} contains $n_{p}=14$
parameters to be determined: $r_{1}$, $r_{2}$, $r_{3}$, $r_{4}$, $\varepsilon
_{11}$, $\varepsilon_{12}$, $\varepsilon_{14}$, $\varepsilon_{21}$, $\varepsilon
_{22}$, $\varepsilon_{23}$, $\varepsilon_{24}$, $\varepsilon_{32}$, $\varepsilon
_{33}$, $\varepsilon_{34}$ (see table \ref{tbl:1} where we explicitly give the elements for
the model of Fig. 5 together with their neighbors). The number of parameters
for the model of Fig. 3, that has $N=108$ elements in each state (see tables \ref{tbl:2a} - \ref{tbl:2f} in the APPENDIX A), is several times greater than this. In Subsection C we
reduce even more the number of parameters for the toy model by equaling some
of the coupling parameters $\varepsilon_{ij}$.%

\subsection{Life and the edge of chaos}

It is generally accepted that the notable stability of the living systems, say
the capacity to support their temporal and spatial organization by adapting
themselves to changes in the environment, is an indication that they evolve at
the edge of chaos\cite{Packard2, Kauffman1,Thurner1}. At the edge of chaos the
system dynamics is characterized by the fact that the largest Lyapunov
exponent is equal to zero. Within a hyperbolic dynamical systems framework,
Pesin, Katok and Ruelle have proved that negative Lyapunov exponents
correspond to global stable manifolds or contracting directions, and positive
Lyapunov exponents correspond to global unstable manifolds or expanding
directions\cite{Pesin1,Katok2,Ruelle1}. In general, a zero exponent
corresponds to a neutral direction, a situation that can be observed in
partially hyperbolic diffeomorphisms\cite{Brin1,Burns1}. For these, the
tangent bundle $TM$ can be split into three invariant continuous subbundles:
$TM=E^{s}\oplus E^{c}\oplus E^{u}$, where $E^{s}$, $E^{u}$ denote the strongly
stable and unstable (respectively) subspaces and $E^{c}$ the central subspace
in which contractions and expansions are weaker. Under certain conditions has
been proved that the largest Lyapunov exponent in this region is
zero\cite{Ponce1}.\ 

Thus, from the dynamical systems theory point of view, we take the largest
Lyapunov exponent equal to zero as being the main characteristic of the living
systems and use this property to determine the unknown parameters in the map
of Eq. \ref{23}.

On the other hand, if we define $DT_{\mathbf{\tilde{B}}}^{k}$ (see Eq.
\ref{18}) as the product of derivatives along the orbit:%

\begin{equation}
DT_{\mathbf{\tilde{B}}}^{k}=DT_{\mathbf{\tilde{B}}_{k-1}}\cdots
DT_{\mathbf{\tilde{B}}_{0}}, \tag{24}\label{24}%
\end{equation}
where $\mathbf{\tilde{B}}_{j}=T^{j}\left(  \mathbf{\tilde{B}}\right)  $, then
Kingman%
\'{}%
s subadditive ergodic theorem\cite{Steele1} says that $\lim_{k\rightarrow
\infty}\frac{1}{k}\log\left(  \left\Vert DT_{\mathbf{\tilde{B}}}%
^{k}\right\Vert \right)  $ exists for almost all points $\mathbf{\tilde{B}}$.
Besides, since\cite{Oseledets1}
\begin{equation}
\underset{k\rightarrow\infty}{\lim\sup}\frac{1}{k}\log\left\Vert
DT_{\mathbf{\tilde{B}}}^{k}\mathbf{v}\right\Vert \leq\lim_{k\rightarrow\infty
}\frac{1}{k}\log\left\Vert DT_{\mathbf{\tilde{B}}}^{k}\right\Vert ,
\tag{25}\label{25}%
\end{equation}
we have for the largest Lyapunov exponent:%
\begin{equation}
\lambda_{\max}\left(  \mathbf{\tilde{B}}\right)  =\lim_{k\rightarrow\infty
}\frac{1}{k}\log\left\Vert DT_{\mathbf{\tilde{B}}}^{k}\right\Vert .
\tag{29}\label{29}%
\end{equation}

Therefore, according to our assumption, the condition of life at the edge of
the chaos would imply that the right hand side of Eq. \ref{29} is equal to zero.

In order to determine the $n_{p\text{ }}$parameters of the map in
Eq.(\ref{23}) we propose to solve the system of non linear equations
$\lambda_{\max}\left(  \mathbf{\tilde{B}}_{\alpha}\right)  =0$ for $n_{i}$
initial states $\mathbf{\tilde{B}}_{\alpha}$ ($\alpha=1,2,\cdots,n_{i}$).
Optimization technics such as genetic algorithms\cite{Goldberg1,Mitchell2} or
the particle swarm optimization procedure\cite{Kennedy1} seem adequate to
numerically find them by using as fitness functions these equations. Of course
the condition $\lambda_{\max}\left(  \mathbf{\tilde{B}}_{\alpha}\right)  =0$
must be interpreted, from a numerical point of view, in the sense of Lyapunov
exponent zero-crossing\cite{Albers2}. We must also mention the possibility
that the diverse initial conditions $\mathbf{\tilde{B}}_{\alpha}$
($\alpha=1,2,\cdots,n_{i}$) evolves towards distinct attractors, so we
consider all the attractors as describing our cell if, for each
$\mathbf{\tilde{B}}_{\alpha}$, the corresponding maximun Lyapunov exponent is zero.

\subsection{Toy model for the cellular fission}

\begin{figure}[t!]
 \centering
 \includegraphics[width=\textwidth]{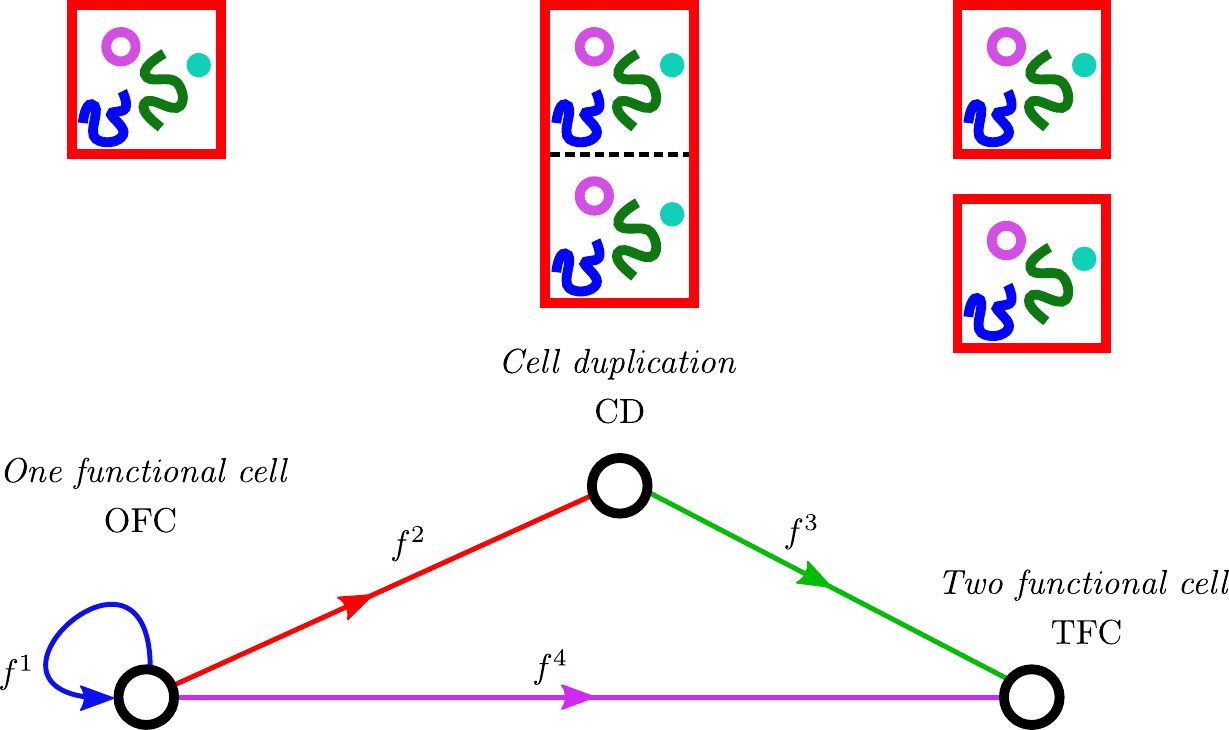}
 \caption{Toy model for the living single cell, with emphasis on the
cellular fission, obtained as a drastic simplification of the more complete
model of Fig. 3.}
 \label{fig:05}
\end{figure}

\begin{table}
\centering
\begin{tabular}{ll}
 \toprule
 $x_{i}$ & neighbors \\
 \midrule
 $x_{1}=\text{OFC}\overset{f_{1}}{\rightarrow}\text{OFC}$ & $x_{1},x_{2},x_{4}$ \\
 $x_{2}=\text{OFC}\overset{f_{2}}{\rightarrow}\text{CD}$ & $x_{1},x_{2},x_{3},x_{4}$ \\
 $x_{3}=\text{CD}\overset{f_{3}}{\rightarrow}\text{TFC}$ & $x_{2},x_{3},x_{4}$ \\
 $x_{4}=\text{OFC}\overset{f_{4}}{\rightarrow}\text{TFC}=x_{3}\bullet x_{2}$ & $x_{1},x_{2},x_{3},x_{4}$ \\
 \bottomrule
\end{tabular}
\caption{Elements of the toy model of Fig. 5 and their neighbors.}
\label{tbl:1}
\end{table}

In Fig. 5 we show the toy model we propose to describe, within our formalism,
the cellular division of a bacterium\cite{Harry1},\cite{Wang1}. It
approximately corresponds to the higher level in the hierarchical category
displayed in Fig. 3 to represent a living single cell. Their elements, with
the corresponding neighbors, are listed in table \ref{tbl:1}.

In the model, the element $x_{1}$= OFC $\overset{f_{1}}{\rightarrow}$ OFC
accounts for the normal workings of a single cell. It includes all the
operations that allow the cell functions along the whole life cycle. The
element $x_{2}$= OFC $\overset{f_{2}}{\rightarrow}$ CD denotes the duplication
of the bacterium chromosome and molecules as well as the mechanisms that
double the bacterium size, all these resulting into the FtsZ ring formation
and the divisome assembly. The element $x_{3}$= CD $\overset{f_{3}%
}{\rightarrow}$ TFC, on the other hand, is associated with the division or
splitting of the living cell into two identical ones. Finally, $x_{4}$= OFC
$\overset{f_{4}}{\rightarrow}$ TFC = $x_{3}\bullet x_{2}$ represents the
composition of both, the cell matter-volume duplication and its separation
into two daughters cells so describing, in a unique process, the
transformation of a single bacterium into a pair of bacteria equal to that one.

At a given time our system is represented as a fuzzy category or, according to
previous comments, as a point $\mathbf{\tilde{A}=}$ $\left(  \mathbf{\chi
}_{\mathbf{A}}(x_{1}),\mathbf{\chi}_{\mathbf{A}}(x_{2}),\mathbf{\chi
}_{\mathbf{A}}(x_{3}),\mathbf{\chi}_{\mathbf{A}}(x_{4})\text{ }\right)  $
$\mathbf{\in}$ $\left[  0,1\right]  ^{4}$ (see Eq.\ref{11}). We remember that
the membership degree $\mathbf{\chi}_{\mathbf{A}}(x_{i})$ denotes the strength
or powerful of the element (activity or process) $x_{i}$ ($i=1,2,3,4$). \ 

The system evolves with time through a sequence of categories driven by the
transformation given by Eq. (\ref{23}) with the $n_{p}=14$ parameters
\[
r_{1},r_{2},r_{3},r_{4},\varepsilon_{11},\varepsilon_{12},\varepsilon
_{14},\varepsilon_{21},\varepsilon_{22},\varepsilon_{23},\varepsilon
_{24},\varepsilon_{32},\varepsilon_{33},\varepsilon_{34}%
\]
taking values such that the system to be at the edge of the chaos so that the
corresponding Lyapunov exponent, evaluated by Eq. \ref{29}, equals zero.
Despite the drastic simplification, with respect to the general model of Fig.
3, implied by the toy model considered here, it even involves a number of
parameters which is large enough as to make the associated dynamics very
complex. In particular, the simultaneous determinations of the $n_{p\text{ }}%
$parameters of the map in Eq.(\ref{23}) is still a hard task. So, here, in
order to exemplify the kind of information we can extract from the model, we
relax the requirement and demand that the condition $\lambda_{\max}\left(
\mathbf{\tilde{B}}_{\alpha}\right)  =0$ be fulfilled for just one initial
condition. Thus we fix all the parameters except one, say $r_{3}$, to some
reasonable values which are suggested by the analysis of the one-dimensional
Ricker map and by the assumption that the membership value of the element
$x_{i}$ at a given time instant, is mainly influenced by its membership value
at the previous iteration and, at less extent, by the membership values of
their neighbors also at the previous time instant. We specifically choose:
$r_{1}=1.0$, $r_{2}=2.5$, $r_{4}=4.6$ . Besides, to simplify even more the
model, we take all the diagonal coupling parameters equal to one and all the
off-diagonal coupling parameters equal to $10^{-3}$: $\varepsilon_{ii}=1$
($i=1,2,3$) and $\varepsilon_{ij}=\varepsilon=0.001$ ($i\neq j$; $i=1,2,3$;
$j$ such that $x_{j}$ is neighbor of $x_{i}$). Finally, the parameter $r_{3}$
will be chosen by requiring that the system maximum Lyapunov exponent equals zero.

\begin{figure}[h]
 \centering
 \includegraphics[width=\textwidth]{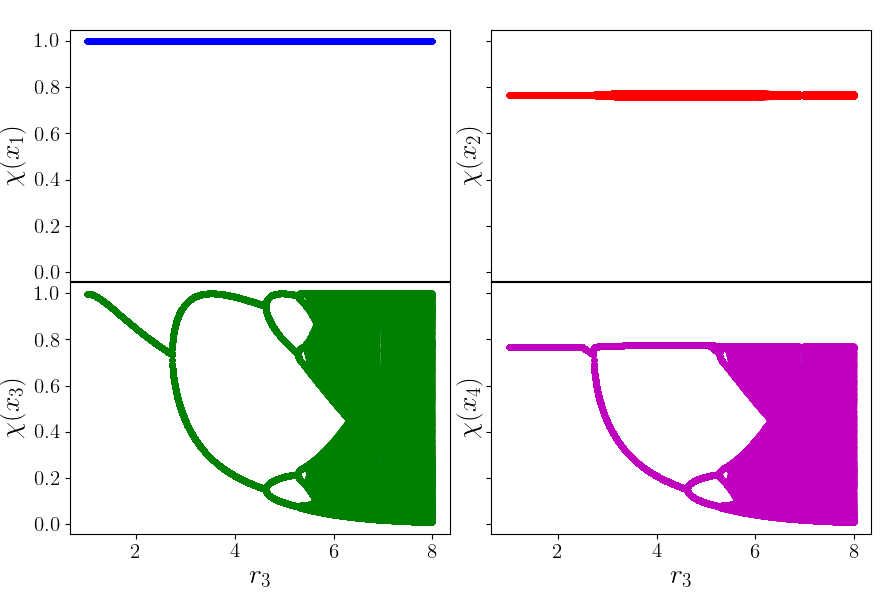}
 \caption{Bifurcation diagrams for $\mathbf{\chi}(x_{i})$
($i=1,2,3,4$)\ as a function of $r_{3}$.}
 \label{fig:06}
\end{figure}

In Fig. 6 we show the bifurcation diagrams for $\mathbf{\chi}(x_{i})$
($i=1,2,3,4$)\ as a function of $r_{3}$ with the remainder parameters fixed at
the indicated values. The corresponding curve for the maximum Lyapunov
exponent is given in Fig. 7.

\begin{figure}[h]
 \centering
 \includegraphics[width=\textwidth]{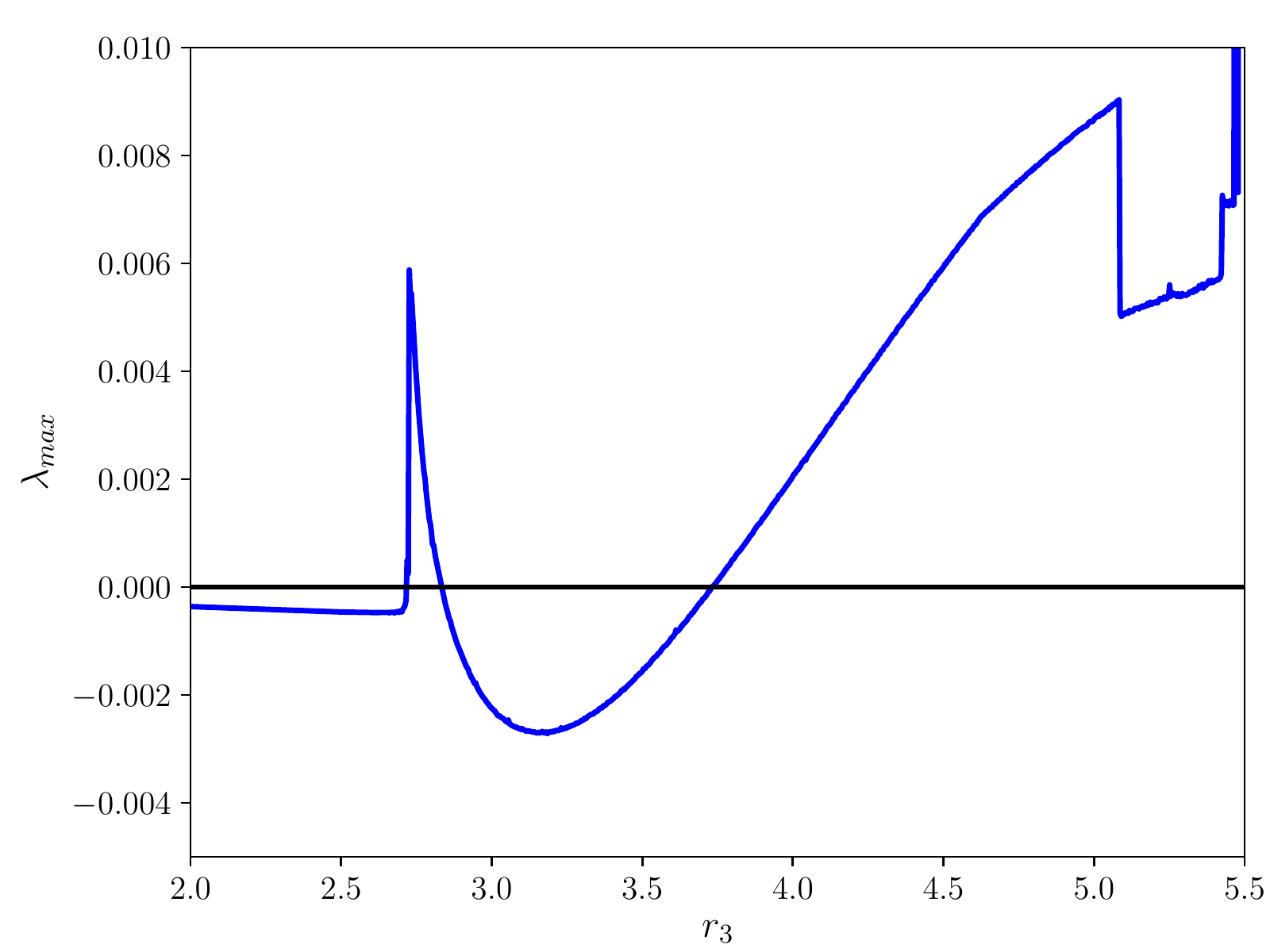}
 \caption{Maximum Lyapunov exponent $\lambda_{\max}$ as a function of
the parameter $r_{3}$.}
 \label{fig:07}
\end{figure}

The bifurcation diagrams for $\mathbf{\chi}(x_{1})$ and $\mathbf{\chi}(x_{2})$
show a period-$2$ cycle along the considered range of $r_{3}$ values, the two
branches being very near one of the other, specially in the case of
$\mathbf{\chi}(x_{1})$. The diagram for $\mathbf{\chi}(x_{3})$ shows a richer
structure with successive bifurcations corresponding to $2^{n}$-cycles
($n=1,2,3,...$). These bifurcations come faster and faster resulting into a
cascade. Taking into account the small values we are considering for the
off-diagonal coupling parameters $\varepsilon_{3j}$ is natural that this
diagram to be very similar to that obtained for the one-dimensional Ricker map
as a function of its parameters $r$ (see APPENDIX B). The diagram for
$\mathbf{\chi}(x_{4})$, on the other hand, is determined by the composition
law between $\mathbf{\chi}(x_{2})$ and $\mathbf{\chi}(x_{3})$.

\begin{figure}[t!]
 \centering
 \includegraphics[scale=0.5]{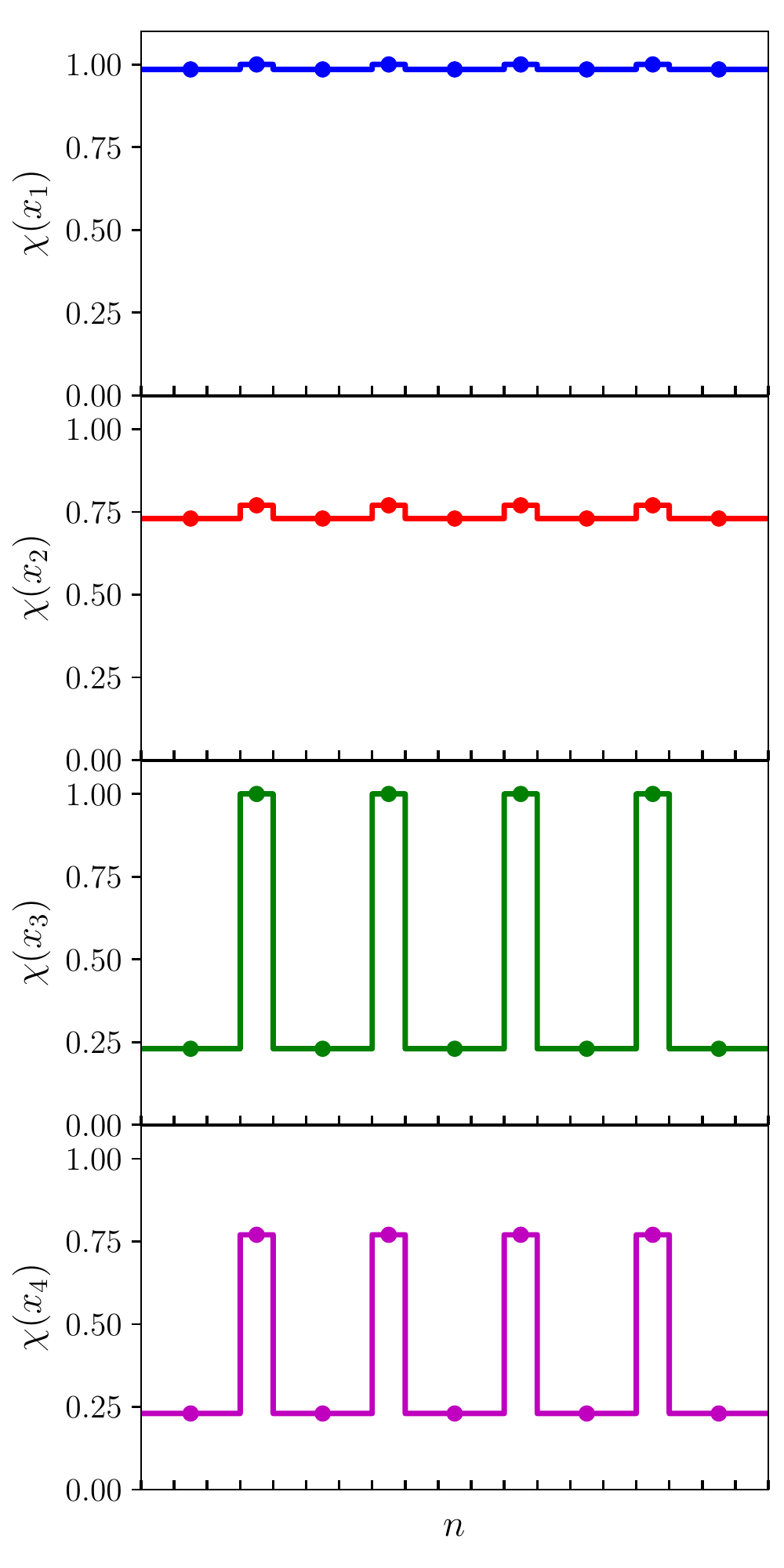}
 \caption{Discrete trajectories $\mathbf{\chi}(x_{i})$ ($i=1,2,3,4$)
vs. $n$ (full circles). Solid lines: continuous representation.}
 \label{fig:08}
\end{figure}

>From the Lyapunov exponent curve (Fig. 7), we take for the parameter $r_{3}$
the value $3.73$. Around this value, $\lambda_{\max}$ changes from being
negative to positive, that is to say the system is at the edge of the chaos.
With $r_{3}=3.73$ and the remainder parameters chosen as indicated before, we
calculate the trajectories $\mathbf{\chi}(x_{i})$ ($i=1,2,3,4$) vs. $n$ ($n$ =
number of iterations)\ determined, in our discrete representation, by the full
circles of Fig. 8. In order to interpret this trajectories in terms of the
biological phenomena in which we are interested, we represent the (continuous)
temporal evolution of the $\mathbf{\chi}(x_{i})%
\acute{}%
s$ with the solid lines. We assume that the interval of time that demands the
proper duplication is several times smaller than the time the cell needs to
prepare itself for that event.

We observe that the four curves are periodic of period $2$. For $\mathbf{\chi
}(x_{1})$ the variation within each period is very small. The same occurs for
$\mathbf{\chi}(x_{2})$ although here the difference is most notable. Actually,
the values $r_{1}=1.0$ and $r_{2}=2.5$ would correspond in the case of a
single Ricker map (see Fig. 9b in the APPENDIX B) to a steady state (fixed
point) at $\mathbf{\chi}(x_{1})\approx1.0$ and $\mathbf{\chi}(x_{2})\approx$
$0.75$, respectively. The small variations, giving period-$2$ curves, are due
to the coupling with the other elements in the generalized map of Eq.
\ref{23}. We interpret this behavior of $\mathbf{\chi}(x_{1})$ as that the
cell works practically at full along its life cycle. Most of the workings is
addressed to prepare the bacterium to his division, so, although to a lesser
strength, $x_{2}$ works almost constant too, increasing a little its power in
the second part of the period. For $r_{3}=3.73$ the bifurcation diagram of the
Fig. 9b in the APPENDIX B gives a period-$2$ cycle with the upper and lower
values at about $0.95$ and $0.24$, respectively which are practically the
values observed in Fig. 8 for $\mathbf{\chi}(x_{3})$, the effect of the
coupling with the other elements being very small. The explanation we can give
for the jump between the first and the second part of the period is that
during the first part the divisome is not yet completely formed and the
division works at full just in the second part. Finally $\mathbf{\chi}(x_{4}%
)$, the curve that globally describe the binary fission, results of the
composition of $\mathbf{\chi}(x_{2})$ and $\mathbf{\chi}(x_{3})$ taking,
between both values, the minimum one so that it oscillates between $0.77$ and
$0.23$.

\section{Conclusions}

In this work we have improved the theory of hierarchical evolutive systems of
Ehresmann and Vandremeersch by adding the concept of fuzzy categories. This
way each category $\mathbf{\tilde{A}}$ can be represented as a point in a
states space $\left[  0,1\right]  ^{N}$ $\subset$\ $%
\mathbb{R}
^{N}$ where $N$ is the space dimension. The coordinates of the points in this
space are the membership values associated with the category components.
Within this context, the state of the system under study at a given time is
described by a hierarchical fuzzy category $\mathbf{\tilde{A}}_{h}$ where the
element $x_{i}=a\overset{f_{i}}{\rightarrow}b$ represents the relation between
the system components $a$ and $b$ mediated by the function $f_{i}$ and
$\mathbf{\chi}_{\mathbf{A}}(x_{i})$ the strength or powerful of this relation
at that time. In our discrete time description, the temporal evolution of the
system is accounted by a sequence of fuzzy categories $\left\{  \left(
\mathbf{\tilde{A}}_{h}\right)  _{n}\right\}  _{n\in%
\mathbb{N}
}$ all whose members have the same elements $x_{i}$ differing in their
membership values $\mathbf{\chi}(x_{i})$ ($i=1,2,\cdots,N$), so that the
functor that transforms the category $\left(  \mathbf{\tilde{A}}_{h}\right)
_{n}$ into the category $\left(  \mathbf{\tilde{A}}_{h}\right)  _{n+1}$ is
determined by simply giving the maps that transform $\mathbf{\chi}_{\left(
\mathbf{A}\right)  _{n}}(x_{i})$ into $\mathbf{\chi}_{\left(  \mathbf{A}%
\right)  _{n+1}}(x_{i})$ say: $\mathbf{\chi}_{\left(  \mathbf{A}\right)
_{n+1}}(x_{i})=T_{i}\left(  \mathbf{\chi}_{\left(  \mathbf{A}\right)  _{n}%
}(x_{i});\text{ }\left\{  \mathbf{\chi}_{\left(  \mathbf{A}\right)  _{n}%
}(x_{j})\right\}  _{j\neq i}\text{ }\right)  $ ($i=1,2,\cdots,N$). If we
consider for the $T_{i}%
\acute{}%
s$ adequate diffeomorphisms, then we can make contact with the theory of
dynamical systems an use all its tools to describe the behavior of the system
with time. In this manner a quantification of the hierarchical evolutive
systems theory is achieved.

Here we have applied our formalism to describe the living single cell for
which we propose a quite general hierarchical category to represent its state
at a given time and have proposed as dynamics parametrized transformations
$T_{i}$ based on the Ricker map. Also we suggest that the parameters be
determined under the condition that the system evolves at the edge of the
chaos, a property generally taken as hallmark of the life. To exemplify the
theory we have drastically simplified the model into a toy model with
dimension $4$ that emphasizes the cellular fission. However we think that the
formalism is applicable to other biological phenomena as well as to diverse
economic and social problems and, in general, to complex systems for which the
ordinary mathematical tools are hard to use. Of course the choice of the maps
to be used as dynamics and the conditions that determine the parameters will
depend on the particular problem to be studied.

\textbf{Acknowledgments}

Support of this work by Universidad Nacional de La Plata, Universidad Nacional
de Rosario and Consejo Nacional de Investigaciones Cient\'{\i}ficas y
T\'{e}cnicas of Argentina is greatly appreciated. C.M.C. and F.V. are a member
and a researcher under contract, respectively, of CONICET.

\newpage

\begin{center}
APPENDIX A: Elements of the living cell model given by Fig. 3
\end{center}

In tables \ref{tbl:2a} - \ref{tbl:2f} we explicitly show the elements involved in the living single cell
model drawn in Fig. 3. To make the presentation more clear we collect in table \ref{tbl:2a} the elements that relate the cell with the environment, in table \ref{tbl:2b} and \ref{tbl:2c}, respectively, the elements of the patterns that determine the
colimits cOFC and cDC, cTFC; in tables \ref{tbl:2d}, \ref{tbl:2e} those involved in the cluster
G$_{\text{8}}$ and clusters G$_{\text{9}}$, G$_{\text{10}}$, repectively, and
finally in table \ref{tbl:2f} the maps linking the colimits cOFC, cDC and cTFC which
accounts for the binary fission.

Elements that differ in a prime symbol (non prime, prime, double prime) have
the same source and target (and thus the same neighbors) but distinct maps.
When it corresponds we have indicated the product of elements given by the
composition law.

\begin{table}[h!]
\centering
\begin{tabular}{ll}
 \toprule
 $x_{i}$ & $x_{i}$ \\
 \midrule
 $x_{1}=\text{Env}\overset{f_{1}}{\rightarrow}\text{M}$ & $x_{2}=\text{Env}\overset{f_{2}}{\rightarrow}\text{E}$ \\
 $x_{1}^{\prime}=\text{Env}\overset{f_{1}^{\prime}}{\rightarrow}\text{M}=x_{6}\bullet x_{2}$ & $x_{2}^{\prime}=\text{Env}\overset{f_{2}^{\prime}}{\rightarrow}\text{E}=x_{5}\bullet x_{1}$ \\
 $x_{3}=\text{M}\overset{f_{3}}{\rightarrow}\text{Env}$ & $x_{4}=\text{E}\overset{f_{4}}{\rightarrow}\text{Env}$ \\
 $x_{3}^{\prime}=\text{M}\overset{f_{3}^{\prime}}{\rightarrow}\text{Env}=x_{4}\bullet x_{5}$ & $x_{4}^{\prime}=\text{E}\overset{f_{4}^{\prime}}{\rightarrow}\text{Env}=x_{3}\bullet x_{6}$ \\
 $x_{5}=\text{M}\overset{f_{5}}{\rightarrow}\text{E}$ & $x_{6}=\text{E}\overset{f_{6}}{\rightarrow}\text{M}$ \\
 $x_{5}^{\prime}=\text{M}\overset{f_{5}^{\prime}}{\rightarrow}\text{E}=x_{2}\bullet x_{3}$ & $x_{6}^{\prime}=\text{E}\overset{f_{6}^{\prime}}{\rightarrow}\text{M}=x_{1}\bullet x_{4}$ \\
 $x_{7}=\text{M}\overset{f_{7}}{\rightarrow}\text{cME}$ & $x_{8}=\text{E}\overset{f_{8}}{\rightarrow}\text{cME}$ \\
 $x_{9}=\text{cPro}\overset{f_{9}}{\rightarrow}\text{Env}$ & \\ 
 \bottomrule
\end{tabular}
\caption{Elements relating the cell and the environment.}
\label{tbl:2a}
\end{table}

\begin{table}
\centering
\begin{tabular}{ll}
 \toprule
 $x_{i}$ & $x_{i}$ \\
 \midrule
 $x_{10}=\text{cME}\overset{f_{10}}{\rightarrow}\text{cMet}$ & $x_{11}=\text{cMet}\overset{f_{11}}{\rightarrow}\text{cME}$ \\
 $x_{12}=\text{cMet}\overset{f_{12}}{\rightarrow}\text{cDNA}$ & $x_{13}=\text{cDNA}\overset{f_{13}}{\rightarrow}\text{cMet}$ \\
 $x_{12}^{\prime}=\text{cMet}\overset{f_{12}^{\prime}}{\rightarrow}\text{cDNA}=x_{19}\bullet x_{14}$ & $x_{13}^{\prime}=\text{cDNA}\overset{f_{13}^{\prime}}{\rightarrow}\text{cMet}=x_{15}\bullet x_{18}$ \\
 $x_{12}^{\prime\prime}=\text{cMet}\overset{f_{12}^{\prime\prime}}{\rightarrow}\text{cDNA}=x_{21}\bullet x_{16}$ & $x_{13}^{\prime\prime}=\text{cDNA}\overset{f_{13}^{\prime\prime}}{\rightarrow}\text{cMet}=x_{17}\bullet x_{20}$ \\
 $x_{14}=\text{cMet}\overset{f_{14}}{\rightarrow}\text{cRNA}$ & $x_{15}=\text{cRNA}\overset{f_{15}}{\rightarrow}\text{cMet}$ \\
 $x_{14}^{\prime}=\text{cMet}\overset{f_{14}^{\prime}}{\rightarrow}\text{cRNA}=x_{18}\bullet x_{12}$ & $x_{15}^{\prime}=\text{cRNA}\overset{f_{15}^{\prime}}{\rightarrow}\text{cMet}=x_{13}\bullet x_{19}$ \\
 $x_{14}^{\prime\prime}=\text{cMet}\overset{f_{14}^{\prime\prime}}{\rightarrow}\text{cRNA}=x_{23}\bullet x_{16}$ & $x_{15}^{\prime\prime}=\text{cRNA}\overset{f_{15}^{\prime\prime}}{\rightarrow}\text{cMet}=x_{17}\bullet x_{22}$ \\
 $x_{16}=\text{cMet}\overset{f_{16}}{\rightarrow}\text{cPro}$ & $x_{17}=\text{cPro}\overset{f_{17}}{\rightarrow}\text{cMet}$ \\
 $x_{16}^{\prime}=\text{cMet}\overset{f_{16}^{\prime}}{\rightarrow}\text{cPro}=x_{22}\bullet x_{14}$ & $x_{17}^{\prime}=\text{cPro}\overset{f_{17}^{\prime}}{\rightarrow}\text{cMet}=x_{15}\bullet x_{23}$ \\
 $x_{16}^{\prime\prime}=\text{cMet}\overset{f_{16}^{\prime\prime}}{\rightarrow}\text{cPro}=x_{20}\bullet x_{12}$ & $x_{17}^{\prime\prime}=\text{cPro}\overset{f_{17}^{\prime\prime}}{\rightarrow}\text{cMet}=x_{13}\bullet x_{21}$ \\
 $x_{18}=\text{cDNA}\overset{f_{18}}{\rightarrow}\text{cRNA}$ & $x_{19}=\text{cRNA}\overset{f_{19}}{\rightarrow}\text{cDNA}$ \\
 $x_{18}^{\prime}=\text{cDNA}\overset{f_{18}^{\prime}}{\rightarrow}\text{cRNA}=x_{14}\bullet x_{13}$ & $x_{19}^{\prime}=\text{cRNA}\overset{f_{19}^{\prime}}{\rightarrow}\text{cDNA}=x_{12}\bullet x_{15}$ \\
 $x_{18}^{\prime\prime}=\text{cDNA}\overset{f_{18}^{\prime\prime}}{\rightarrow}\text{cRNA}=x_{23}\bullet x_{20}$ & $x_{19}^{\prime\prime}=\text{cRNA}\overset{f_{19}^{\prime\prime}}{\rightarrow}\text{cDNA}=x_{21}\bullet x_{22}$ \\
 $x_{20}=\text{cDNA}\overset{f_{20}}{\rightarrow}\text{cPro}$ & $x_{21}=\text{cPro}\overset{f_{21}}{\rightarrow}\text{cDNA}$ \\
 $x_{20}^{\prime}=\text{cDNA}\overset{f_{20}^{\prime}}{\rightarrow}\text{cPro}=x_{16}\bullet x_{13}$ & $x_{21}^{\prime}=\text{cPro}\overset{f_{21}^{\prime}}{\rightarrow}\text{cDNA}=x_{12}\bullet x_{17}$ \\
 $x_{20}^{\prime\prime}=\text{cDNA}\overset{f_{20}^{\prime\prime}}{\rightarrow}\text{cPro}=x_{22}\bullet x_{18}$ & $x_{21}^{\prime\prime}=\text{cPro}\overset{f_{21}^{\prime\prime}}{\rightarrow}\text{cDNA}=x_{19}\bullet x_{23}$ \\
 $x_{22}=\text{cRNA}\overset{f_{22}}{\rightarrow}\text{cPro}$ & $x_{23}=\text{cPro}\overset{f_{23}}{\rightarrow}\text{cRNA}$ \\
 $x_{22}^{\prime}=\text{cRNA}\overset{f_{22}^{\prime}}{\rightarrow}\text{cPro}=x_{16}\bullet x_{15}$ & $\text{ }x_{23}^{\prime}=\text{cPro}\overset{f_{23}^{\prime}}{\rightarrow}\text{cRNA}=x_{14}\bullet x_{17}$ \\
 $x_{22}^{\prime\prime}=\text{cRNA}\overset{f_{22}^{\prime\prime}}{\rightarrow}\text{cPro}=x_{20}\bullet x_{21}$ & $x_{23}^{\prime\prime}=\text{cPro}\overset{f_{23}^{\prime\prime}}{\rightarrow}\text{cRNA}=x_{18}\bullet x_{21}$ \\
 $x_{24}=\text{cME}\overset{f_{24}}{\rightarrow}\text{cME}$ & $x_{25}=\text{cMet}\overset{f_{25}}{\rightarrow}\text{cMet}$ \\ 
 $x_{26}=\text{cDNA}\overset{f_{26}}{\rightarrow}\text{cDNA}$ & $x_{27}=\text{cRNA}\overset{f_{27}}{\rightarrow}\text{cRNA}$ \\
 $x_{28}=\text{cPro}\overset{f_{28}}{\rightarrow}\text{cPro}$ & $x_{29}=\text{cME}\overset{f_{29}}{\rightarrow}\text{cOFC}$ \\
 $x_{30}=\text{cMet}\overset{f_{30}}{\rightarrow}\text{cOFC}$ & $\text{ }x_{31}=\text{cDNA}\overset{f_{31}}{\rightarrow}\text{cOFC}$ \\
 $x_{32}=\text{cRNA}\overset{f_{32}}{\rightarrow}\text{cOFC}$ & $x_{33}=\text{cPro}\overset{f_{33}}{\rightarrow}\text{cOFC}$ \\
 \bottomrule
\end{tabular}
\caption{Elements of the patterns determining cOFC.}
\label{tbl:2b}
\end{table}

\begin{table}
\centering
\begin{tabular}{ll}
 \toprule
 $x_{i}$ & $x_{i}$ \\
 \midrule
 $x_{34}=\text{MD}\overset{f_{34}}{\rightarrow}\text{CD}$ & $x_{35}=\text{VD}\overset{f_{35}}{\rightarrow}\text{CD}$ \\
 $x_{36}=\text{CD}\overset{f_{36}}{\rightarrow}\text{cDC}$ & $x_{37}=\text{MD}\overset{f_{37}}{\rightarrow}\text{cDC}=x_{36}\bullet x_{34}$ \\
 $x_{38}=\text{VD}\overset{f_{38}}{\rightarrow}\text{cDC}=x_{36}\bullet x_{35}$ & $x_{39}=\text{FC1}\overset{f_{39}}{\rightarrow}\text{FC1}$ \\
 $x_{40}=\text{FC2}\overset{f_{40}}{\rightarrow}\text{FC2}$ & $x_{41}=\text{FC1}\overset{f_{41}}{\rightarrow}\text{cTFC}$ \\
 $x_{42}=\text{FC2}\overset{f_{42}}{\rightarrow}\text{cTFC}$ & \\
 \bottomrule
\end{tabular}
\caption{Elements of the patterns determining cDC and cTFC.}
\label{tbl:2c}
\end{table}

\begin{table}
\centering
\begin{tabular}{ll}
 \toprule
 $x_{i}$ & $x_{i}$ \\
 \midrule
 $x_{43}=\text{cME}\overset{f_{43}}{\rightarrow}\text{MD}$ & $x_{44}=\text{cME}\overset{f_{44}}{\rightarrow}\text{VD}$ \\
 $x_{45}=\text{cME}\overset{f_{45}}{\rightarrow}\text{CD}$ & $x_{46}=\text{cMet} \overset{f_{46}}{\rightarrow}\text{MD}$ \\
 $x_{47}=\text{cMet}\overset{f_{47}}{\rightarrow}\text{VD}$ & $x_{48}=\text{cMet}\overset{f_{48}}{\rightarrow}\text{CD}$ \\
 $x_{49}=\text{cDNA}\overset{f_{49}}{\rightarrow}\text{MD}$ & $x_{50}=\text{cDNA}\overset{f_{50}}{\rightarrow}\text{VD}$ \\
 $x_{51}=\text{cDNA}\overset{f_{51}}{\rightarrow}\text{CD}$ & $x_{52}=\text{cRNA}\overset{f_{52}}{\rightarrow}\text{MD}$  \\
 $x_{53}=\text{cRNA}\overset{f_{53}}{\rightarrow}\text{VD}$ & $x_{54}=\text{cRNA}\overset{f_{54}}{\rightarrow}$\\
 $x_{55}=\text{cPro}\overset{f_{55}}{\rightarrow}\text{MD}$ & $x_{56}=\text{cPro}\overset{f_{56}}{\rightarrow}\text{VD}$ \\
 $x_{57}=\text{cPro}\overset{f_{57}}{\rightarrow}\text{CD}$ & \\
 \bottomrule
\end{tabular}
\caption{Elements of the clusters \textbf{G}$_{\text{8}}$.}
\label{tbl:2d}
\end{table}

\begin{table}
\centering
\begin{tabular}{ll}
 \toprule
 $x_{i}$ & $x_{i}$ \\
 \midrule
 $x_{58}=\text{MD}\overset{f_{58}}{\rightarrow}\text{FC1}$ & $x_{59}=\text{MD}\overset{f_{59}}{\rightarrow}\text{FC2}$ \\
 $x_{60}=\text{VD}\overset{f_{60}}{\rightarrow}\text{FC1}$ & $x_{61}=\text{VD}\overset{f_{61}}{\rightarrow}\text{FC2}$ \\
 $x_{62}=\text{CD}\overset{f_{62}}{\rightarrow}\text{FC1}$ & $4x_{63}=\text{CD}\overset{f_{63}}{\rightarrow}\text{FC}$ \\
 $x_{64}=\text{cME}\overset{f_{64}}{\rightarrow}\text{FC1}$ & $x_{65}=\text{cME}\overset{f_{65}}{\rightarrow}\text{FC2}$ \\
 $x_{66}=\text{cMet}\overset{f_{66}}{\rightarrow}\text{FC1}$ & $x_{67}=\text{cMet}\overset{f_{67}}{\rightarrow}\text{FC2}$ \\
 $x_{68}=\text{cDNA}\overset{f_{68}}{\rightarrow}\text{FC1}$ & $x_{69}=\text{cDNA}\overset{f_{69}}{\rightarrow}\text{FC2}$ \\
 $x_{70}=\text{cRNA}\overset{f_{70}}{\rightarrow}\text{FC1}$ & $x_{71}=\text{cRNA}\overset{f_{71}}{\rightarrow}\text{FC2}$ \\
 $x_{72}=\text{cPro}\overset{f_{72}}{\rightarrow}\text{FC1}$ & $x_{73}=\text{cPro}\overset{f_{73}}{\rightarrow}\text{FC2}$ \\
 \bottomrule
\end{tabular}
\caption{Elements of the clusters \textbf{G}$_{\text{9}}$ and \textbf{G}$_{\text{10}}$.}
\label{tbl:2e}
\end{table}

\begin{table}
\centering
\begin{tabular}{ll}
 \toprule
 $x_{i}$ & $x_{i}$ \\
 \midrule
 $x_{74}=\text{cOFC}\overset{f_{74}}{\rightarrow}\text{cDC}$ & $x_{75}=\text{cDC}\overset{f_{75}}{\rightarrow}\text{cTFC}$ \\
 $x_{76}=\text{cOFC}\overset{f_{76}}{\rightarrow}\text{cTFC}$ & $x_{77}=\text{cTFC}\overset{f_{77}}{\rightarrow}\text{cOFC}$ \\
 $x_{78}=\text{cTFC}\overset{f_{78}}{\rightarrow}\text{Env}$ & \\
 \bottomrule
\end{tabular}
\caption{Elements accounting for the binary fission.}
\label{tbl:2f}
\end{table}

\begin{center}
\newpage APPENDIX B: The Ricker map
\end{center}

In Fig. 9(a) we show the Ricker map normalized to transform the interval $\left[
0,1\right]  $ into itself, say
\[
x_{n+1}=rx_{n}\exp\left(  1-rx_{n}\right)  ,
\]
with the parameter $r$ taking the values used in the text. Fig. 9(b), on the
other hand, shows the corresponding bifurcation diagram for $r\in\left[
1,8\right]  $.%

\begin{figure}[h]
 \centering
 \begin{subfigure}[c]{0.6\textwidth}
 \includegraphics[width=\textwidth]{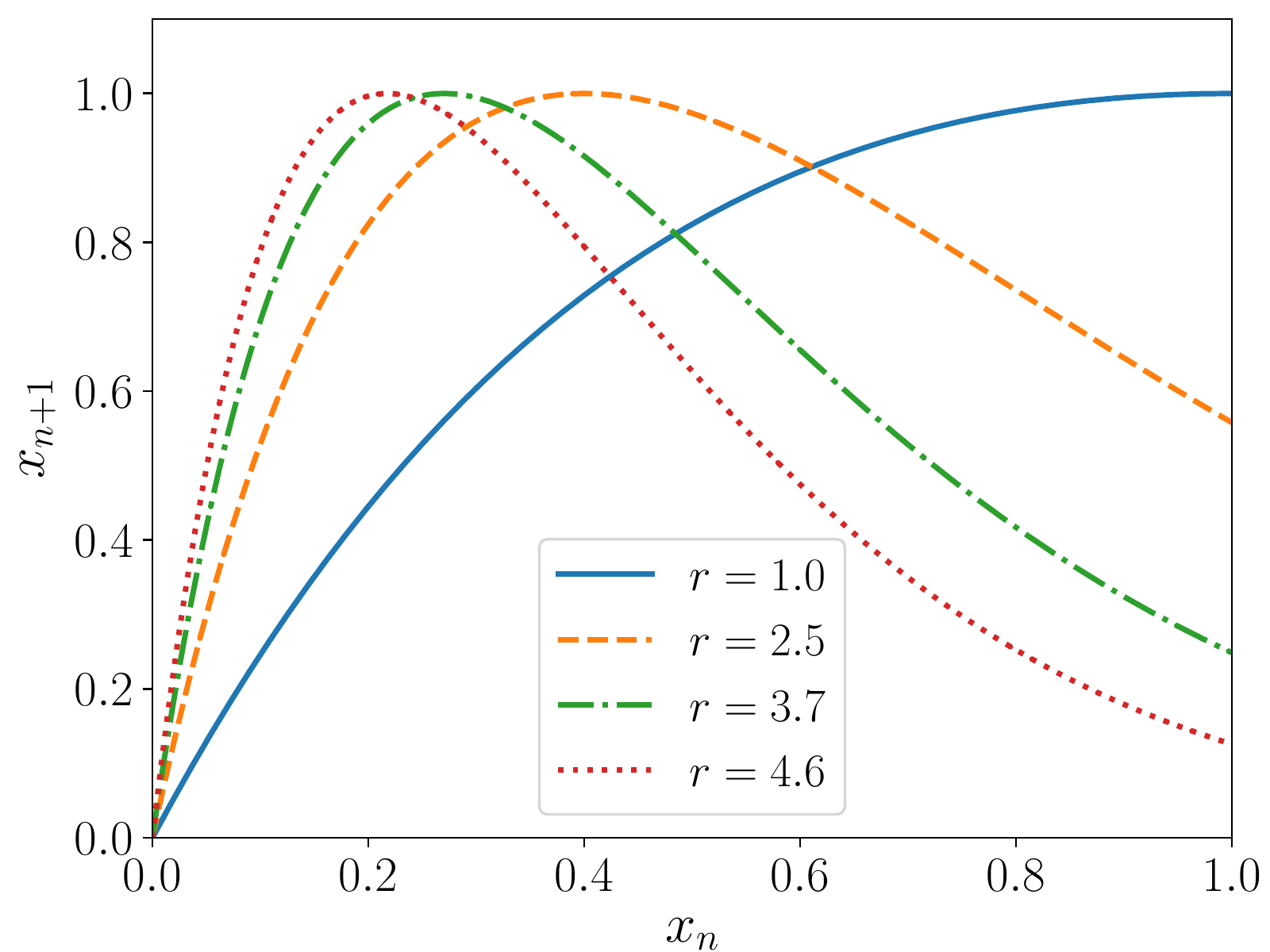}
 \caption{}
 \end{subfigure}
 \begin{subfigure}[c]{0.6\textwidth}
  \centering
  \includegraphics[width=\textwidth]{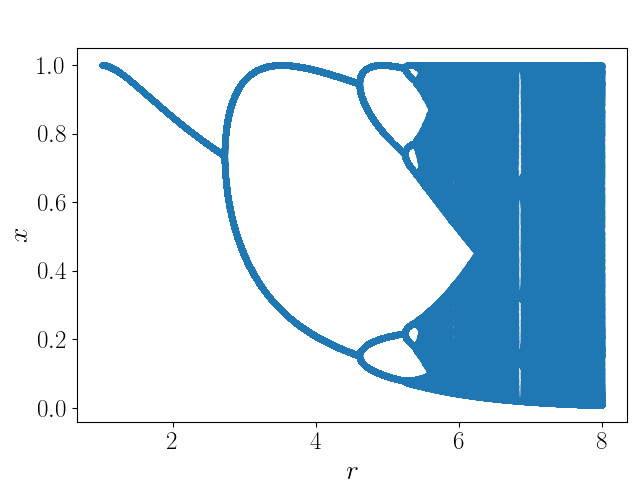}
  \caption{}
 \end{subfigure}
 \caption{Panel\textbf{ }a: Discrete trajectories $\mathbf{\chi
}(x_{i})$ ($i=1,2,3,4$) vs. $n$ (full circles). Solid lines: continuous
representation. Panel b: Bifurcation diagram for a single normalized Ricker map.}
 \label{fig:09}
\end{figure}

\section*{References}

\bibliography{references}

\begin{thebibliography}{10}
\expandafter\ifx\csname url\endcsname\relax
  \def\url#1{\texttt{#1}}\fi
\expandafter\ifx\csname urlprefix\endcsname\relax\def\urlprefix{URL }\fi
\expandafter\ifx\csname href\endcsname\relax
  \def\href#1#2{#2} \def\path#1{#1}\fi

\bibitem{Badii1}
R.~Badii, A.~Politi, Complexity. Cambridge Nonlinear Science Series 6,
  Cambridge, UK: Cambridge University Press, 1997.

\bibitem{Gallavotti1}
G.~Gallavotti, Statistical mechanics, Springer-Verlag, Berlin Heidelberg, 1999.

\bibitem{Honerkamp1}
J.~Honerkamp, Estimation of Parameters, Springer Berlin Heidelberg, Berlin,
  Heidelberg, 2002, pp. 309--337.
\newblock \href {http://dx.doi.org/10.1007/978-3-662-04763-7_8}
  {\path{doi:10.1007/978-3-662-04763-7_8}}.

\bibitem{Packard1}
N.~H. Packard, {Complexity of growing patterns in cellular automata}, Tech.
  rep., Princeton Univ. Inst. Adv. Stud., Princeton, NJ (Jan 1984).

\bibitem{Mitchell1}
M.~Mitchell, J.~P. Crutchfield, P.~T. Hraber, Evolving cellular automata to
  perform computations: mechanisms and impediments, Physica D: Nonlinear
  Phenomena 75~(1) (1994) 361 -- 391.
\newblock \href
  {http://dx.doi.org/https://doi.org/10.1016/0167-2789(94)90293-3}
  {\path{doi:https://doi.org/10.1016/0167-2789(94)90293-3}}.

\bibitem{Dorfman1}
J.~R. Dorfman, An Introduction to Chaos in Nonequilibrium Statistical Mechanics
  (Cambridge Lecture Notes in Physics), Cambridge University Press, 1999.

\bibitem{Bertalanffy1}
L.~von Bertalanffy, General system theory: Foundations, development,
  applications (Penguin university books), Penguin, 1973.

\bibitem{Rashevsky1}
N.~Rashevsky, Topology and life: In search of general mathematical principles
  in biology and sociology, The bulletin of mathematical biophysics 16~(4)
  (1954) 317--348.
\newblock \href {http://dx.doi.org/10.1007/BF02484495}
  {\path{doi:10.1007/BF02484495}}.

\bibitem{Rosen1}
R.~Rosen, A relational theory of biological systems, The bulletin of
  mathematical biophysics 20~(3) (1958) 245--260.
\newblock \href {http://dx.doi.org/10.1007/BF02478302}
  {\path{doi:10.1007/BF02478302}}.

\bibitem{Rosen2}
R.~Rosen, A relational theory of biological systems ii, The bulletin of
  mathematical biophysics 21~(2) (1959) 109--128.
\newblock \href {http://dx.doi.org/10.1007/BF02476354}
  {\path{doi:10.1007/BF02476354}}.

\bibitem{Eilenberg1}
S.~Eilenberg, S.~MacLane, General theory of natural equivalences, Transactions
  of the American Mathematical Society 58~(2) (1945) 231--294.

\bibitem{Ehresmann1}
A.~Ehresmann, J.-P. Vanbremeersch, Hierarchical evolutive systems: A
  mathematical model for complex systems, Bulletin of Mathematical Biology
  49~(1) (1987) 13 -- 50.
\newblock \href
  {http://dx.doi.org/https://doi.org/10.1016/S0092-8240(87)80033-2}
  {\path{doi:https://doi.org/10.1016/S0092-8240(87)80033-2}}.

\bibitem{Ehresmann2}
A.~Ehresmann, J.~Vanbremeersch, Memory Evolutive Systems; Hierarchy, Emergence,
  Cognition, Volume 4 (Studies in Multidisciplinarity), Elsevier Science, 2007.

\bibitem{Katok1}
A.~Katok, B.~Hasselblatt, Introduction to the Modern Theory of Dynamical
  Systems (Encyclopedia of Mathematics and its Applications), Cambridge
  University Press, 1996.

\bibitem{Behrisch1}
M.~Behrisch, S.~Kerkhoff, R.~P{\"o}schel, F.~M. Schneider, S.~Siegmund,
  Dynamical systems in categories, Applied Categorical Structures 25~(1) (2017)
  29--57.
\newblock \href {http://dx.doi.org/10.1007/s10485-015-9409-8}
  {\path{doi:10.1007/s10485-015-9409-8}}.

\bibitem{Kontsevich1}
G.~Dimitrov, F.~Haiden, L.~Katzarkov, M.~Kontsevich, Dynamical systems and
  categories, arXiv preprint arXiv:1307.8418 [math.CT].

\bibitem{Zeigler1}
B.~P. Zeigler, R.~Weinberg, System theoretic analysis of models: Computer
  simulation of a living cell, Journal of Theoretical Biology 29~(1) (1970) 35
  -- 56.
\newblock \href
  {http://dx.doi.org/https://doi.org/10.1016/0022-5193(70)90117-7}
  {\path{doi:https://doi.org/10.1016/0022-5193(70)90117-7}}.

\bibitem{Shuler1}
M.~M. Domach, S.~K. Leung, R.~E. Cahn, G.~G. Cocks, M.~L. Shuler, Computer
  model for glucose-limited growth of a single cell of escherichia coli b/r-a,
  Biotechnology and Bioengineering 26~(3) (1984) 203--216.
\newblock \href {http://dx.doi.org/10.1002/bit.260260303}
  {\path{doi:10.1002/bit.260260303}}.

\bibitem{Shuler2}
E.~V. Nikolaev, J.~C. Atlas, M.~L. Shuler, Computer models of bacterial cells:
  from generalized coarsegrained to genome-specific modular models, Journal of
  Physics: Conference Series 46~(1) (2006) 322.

\bibitem{Hartwell1}
L.~H. Hartwell, J.~J. Hopfield, S.~Leibler, A.~W. Murray, From molecular to
  modular cell biology, Nature 402~(6761supp) (1999) C47--C52.
\newblock \href {http://dx.doi.org/10.1038/35011540}
  {\path{doi:10.1038/35011540}}.

\bibitem{Covert1}
M.~W. Covert, N.~Xiao, T.~J. Chen, J.~R. Karr, Integrating metabolic,
  transcriptional regulatory and signal transduction models in escherichia
  coli, Bioinformatics 24~(18) (2008) 2044--2050.
\newblock \href
  {http://arxiv.org/abs//oup/backfile/content_public/journal/bioinformatics/24/18/10.1093_bioinformatics_btn352/2/btn352.pdf}
  {\path{arXiv:/oup/backfile/content_public/journal/bioinformatics/24/18/10.1093_bioinformatics_btn352/2/btn352.pdf}},
  \href {http://dx.doi.org/10.1093/bioinformatics/btn352}
  {\path{doi:10.1093/bioinformatics/btn352}}.

\bibitem{Covert2}
J.~Karr, J.~Sanghvi, D.~Macklin, M.~Gutschow, J.~Jacobs, B.~Bolival,
  N.~Assad-Garcia, J.~Glass, M.~Covert, A whole-cell computational model
  predicts phenotype from genotype, Cell 150~(2) (2012) 389 -- 401.
\newblock \href {http://dx.doi.org/https://doi.org/10.1016/j.cell.2012.05.044}
  {\path{doi:https://doi.org/10.1016/j.cell.2012.05.044}}.

\bibitem{Covert3}
M.~W. Covert, Simulating a living cell., Scientific American 310~(1) (2014) 44.

\bibitem{Anderson1}
P.~W. Anderson, More is different, Science 177~(4047) (1972) 393--396.
\newblock \href
  {http://arxiv.org/abs/http://science.sciencemag.org/content/177/4047/393.full.pdf}
  {\path{arXiv:http://science.sciencemag.org/content/177/4047/393.full.pdf}},
  \href {http://dx.doi.org/10.1126/science.177.4047.393}
  {\path{doi:10.1126/science.177.4047.393}}.

\bibitem{Ricker1}
W.~E. Ricker, Stock and recruitment, Journal of the Fisheries Research Board of
  Canada 11~(5) (1954) 559--623.
\newblock \href {http://arxiv.org/abs/https://doi.org/10.1139/f54-039}
  {\path{arXiv:https://doi.org/10.1139/f54-039}}, \href
  {http://dx.doi.org/10.1139/f54-039} {\path{doi:10.1139/f54-039}}.

\bibitem{Packard2}
N.~H. Packard, Adaptation toward the edge of chaos, Dynamic patterns in complex
  systems 212 (1988) 293--301.

\bibitem{Kauffman1}
S.~A. Kauffman, The Origins of Order: Self-Organization and Selection in
  Evolution, Oxford University Press, 1993.

\bibitem{Thurner1}
R.~Hanel, M.~P{\"o}chacker, S.~Thurner, Living on the edge of chaos: minimally
  nonlinear models of genetic regulatory dynamics, Philosophical Transactions
  of the Royal Society of London A: Mathematical, Physical and Engineering
  Sciences 368~(1933) (2010) 5583--5596.
\newblock \href
  {http://arxiv.org/abs/http://rsta.royalsocietypublishing.org/content/368/1933/5583.full.pdf}
  {\path{arXiv:http://rsta.royalsocietypublishing.org/content/368/1933/5583.full.pdf}},
  \href {http://dx.doi.org/10.1098/rsta.2010.0267}
  {\path{doi:10.1098/rsta.2010.0267}}.

\bibitem{Tsallis1}
C.~Tsallis, A.~Plastino, W.-M. Zheng, Power-law sensitivity to initial
  conditions—new entropic representation, Chaos, Solitons \& Fractals 8~(6)
  (1997) 885 -- 891.
\newblock \href
  {http://dx.doi.org/https://doi.org/10.1016/S0960-0779(96)00167-1}
  {\path{doi:https://doi.org/10.1016/S0960-0779(96)00167-1}}.

\bibitem{Latora1}
V.~Latora, M.~Baranger, A.~Rapisarda, C.~Tsallis, The rate of entropy increase
  at the edge of chaos, Physics Letters A 273~(1) (2000) 97 -- 103.
\newblock \href
  {http://dx.doi.org/https://doi.org/10.1016/S0375-9601(00)00484-9}
  {\path{doi:https://doi.org/10.1016/S0375-9601(00)00484-9}}.

\bibitem{Robledo1}
A.~Robledo, Generalized statistical mechanics at the onset of chaos, Entropy
  15~(12) (2013) 5178--5222.
\newblock \href {http://dx.doi.org/10.3390/e15125178}
  {\path{doi:10.3390/e15125178}}.

\bibitem{Zadeh1}
L.~A. Zadeh, Information and control, Fuzzy sets 8~(3) (1965) 338--353.

\bibitem{Robinson1}
C.~R. Robinson, Dynamical systems, CRC press, 1999.

\bibitem{Oseledets1}
I.~Oseledets, A multiplicative ergodic theorem: Lyapunov characteristic numbers
  for dynamical systems, Transactions of the Moscow Mathematical Society 19
  (1968) 197--231.

\bibitem{Strogatz1}
S.~Strogatz, Nonlinear Dynamics and Chaos, Westview Press, Cambridge,
  Massachusetts, 2000.

\bibitem{Albers1}
D.~Albers, J.~Sprott, Routes to chaos in high-dimensional dynamical systems: A
  qualitative numerical study, Physica D: Nonlinear Phenomena 223~(2) (2006)
  194 -- 207.
\newblock \href {http://dx.doi.org/https://doi.org/10.1016/j.physd.2006.09.004}
  {\path{doi:https://doi.org/10.1016/j.physd.2006.09.004}}.

\bibitem{Pesin1}
Y.~B. Pesin, Characteristic lyapunov exponents and smooth ergodic theory,
  Russian Mathematical Surveys 32~(4) (1977) 55--114.

\bibitem{Katok2}
A.~Katok, Lyapunov exponents, entropy and periodic orbits for diffeomorphisms,
  Inst. Hautes \'{E}tudes Sci. Publ. Math 51~(1) (1980) 137--173.

\bibitem{Ruelle1}
D.~Ruelle, Characteristic exponents and invariant manifolds in hilbert space,
  Annals of Mathematics 115~(2) (1982) 243--290.

\bibitem{Brin1}
M.~I. Brin, J.~B. Pesin, Partially hyperbolic dynamical systems, Mathematics of
  the USSR-Izvestiya 8~(1) (1974) 177.

\bibitem{Burns1}
K.~Burns, D.~Dolgopyat, Y.~Pesin, Partial hyperbolicity, lyapunov exponents and
  stable ergodicity, Journal of statistical physics 108~(5-6) (2002) 927--942.

\bibitem{Ponce1}
G.~Ponce, A.~Tahzibi, Central lyapunov exponent of partially hyperbolic
  diffeomorphisms of 𝕋$^3$, Proceedings of the American Mathematical Society
  142~(9) (2014) 3193--3205.

\bibitem{Steele1}
J.~M. Steele, Kingman’s subadditive ergodic theorem, Ann. Inst. H.
  Poincar{\'e} Probab. Statist 25~(1) (1989) 93--98.

\bibitem{Goldberg1}
D.~E. Goldberg, Genetic Algorithms in Search, Optimization, and Machine
  Learning, Addison-Wesley Professional, 1989.

\bibitem{Mitchell2}
M.~Mitchell, An Introduction to Genetic Algorithms (Complex Adaptive Systems),
  The MIT Press, 1996.

\bibitem{Kennedy1}
J.~Kennedy, R.~C. Eberhart, Particle swarm optimization, in: Proceedings of the
  1995 IEEE International Conference on Neural Networks, Vol.~4, Perth,
  Australia, IEEE Service Center, Piscataway, NJ, 1995, pp. 1942--1948.

\bibitem{Albers2}
D.~J. Albers, J.~C. Sprott,
  \href{http://stacks.iop.org/0951-7715/19/i=8/a=005}{Structural stability and
  hyperbolicity violation in high-dimensional dynamical systems}, Nonlinearity
  19~(8) (2006) 1801.
\newline\urlprefix\url{http://stacks.iop.org/0951-7715/19/i=8/a=005}

\bibitem{Harry1}
E.~Harry, L.~Monahan, L.~Thompson, Bacterial cell division: The mechanism and
  its precison, Vol. 253 of International Review of Cytology, Academic Press,
  2006, pp. 27 -- 94.
\newblock \href
  {http://dx.doi.org/https://doi.org/10.1016/S0074-7696(06)53002-5}
  {\path{doi:https://doi.org/10.1016/S0074-7696(06)53002-5}}.

\bibitem{Wang1}
J.~D. Wang, P.~A. Levin, Metabolism, cell growth and the bacterial cell cycle,
  Nature Reviews Microbiology 7~(11) (2009) 822--827.

\end{thebibliography}

\end{document}